\documentclass[apps,pra,onecolumn,armsfonts,amssymb,armsmath,showpacs,floatfix,nofootinbib,groupedaddress,superscriptaddress,citesort]{revtex4}
\usepackage{mathrsfs}
\usepackage{amsfonts}
\usepackage{amstext}
\usepackage{amsmath}
\usepackage{amssymb}
\usepackage{float}
\usepackage[usenames]{xcolor}
\usepackage[dvips]{graphicx}
\usepackage[colorlinks, citecolor=blue]{hyperref}

\def\ra{\rangle}
\def\la{\langle}

\def\no{\nonumber}
\def\bea{\begin{eqnarray}}
\def\eea{\end{eqnarray}}
\def\be{\begin{equation}}
\def\ee{\end{equation}}

\def\ra{\rangle}
\def\la{\langle}

\def\no{\nonumber}
\def\bea{\begin{eqnarray}}
\def\eea{\end{eqnarray}}
\def\be{\begin{equation}}
\def\ee{\end{equation}}

\begin{document}
\title{Probing Planckian physics in de Sitter space with quantum correlations}

\author{Jun Feng}
\affiliation{School of Mathematics and Physics, The University of Queensland, Brisbane, QLD 4072, Australia}
\author{Yao-Zhong Zhang}
\affiliation{School of Mathematics and Physics, The University of Queensland, Brisbane, Qld 4072, Australia}
\author{Mark D. Gould}
\affiliation{School of Mathematics and Physics, The University of Queensland, Brisbane, Qld 4072, Australia}
\author{Heng Fan}
\affiliation{Institute of Physics, Chinese Academy of Sciences, Beijing 100190, P. R. China}
\author{Cheng-Yi Sun}
\affiliation{Institute of Modern Physics, Northwest University, Xian 710069, P. R. China}
\author{Wen-Li Yang}
\affiliation{Institute of Modern Physics, Northwest University, Xian 710069, P. R. China}


\begin{abstract}

We study the quantum correlation and quantum communication channel of both free scalar and fermionic fields in de Sitter space, while the Planckian modification presented by the choice of a particular $\alpha$-vacuum has been considered. We show that the occurrence of degradation of quantum entanglement between field modes for an inertial observer in curved space, due to the radiation associated with its cosmological horizon. Comparing with standard Bunch-Davies choice, the possible Planckian physics causes some extra decrement on the quantum correlation, which may provide the means to detect quantum gravitational effects via quantum information methodology in future. Beyond single-mode approximation, we construct proper Unruh modes admitting general $\alpha-$vacua, and find a convergent feature of both bosonic and fermionic entanglements. In particular, we show that the convergent points of fermionic entanglement negativity are dependent on the choice of $\alpha$. Moreover, an one-to-one correspondence between convergent points $H_c$ of negativity and zeros of quantum capacity of quantum channels in de Sitter space has been proved.


\end{abstract}

\maketitle

\section{Introduction}

Quantum information processing, based on principles of quantum mechanics, promises algorithms surpassing their classical counterparts and unconditional secure for quantum communication. Combined with relativity which is another cornerstone of modern physics, a new fast-growing field called Relativistic Quantum Information (RQI) has attracted much attentions in recent years (see Ref.\cite{RQI1} for a recent review). Besides its importance in probing extremely sensitive gravitational effects \cite{RQI2}, it can also shed lights on study of black hole information paradox \cite{RQI3} and cosmological evolution \cite{RQI4}. Experimental realizations of RQI process in laboratory systems like atomic interferometry \cite{RQI5} and circuit QED \cite{RQI6} have also been proposed. 

A novel phenomenon in RQI is that quantum correlations of a mostly entangled physical system are highly observer-dependent. For a bipartite entangled system in flat space, it was shown \cite{RQI7} that an accelerated observer, having no access to information beyond his acceleration horizon, would experience some decrement of the quantum correlations he shared initially with an inertial partner. Therefore the information-loss results in the celebrated Unruh effect \cite{UNRUH} which claims the detection of a thermal bath for an accelerated detector in flat space. Such decoherence phenomena should exist for any other kinds of causal horizons related with thermal radiation, e.g., the event horizon of black hole \cite{RQI8}. By a process akin to teleportation but without the classical information transmitted \cite{RQI3,RQI9}, quantum information can even escape from a black hole. While above studies focus on entangled global modes, recently, entanglement between fields or systems localized in spacetime has also attracted much attention \cite{RQI10,RQI11,RQI12,FENG1,FENG2}. These entangled localized systems can in principle be measured and exploited for real quantum information tasks \cite{RQI13}. 

As an idealization of inflation regime in cosmology, de Sitter space possesses a cosmological event horizon, from which a thermal spectrum with temperature $T=H/2\pi$ could be detected by a static observer \cite{GH}. This so-called Gibbons-Hawking effect results from the highly non-trivial definition of a vacuum state in a time-dependent background where no global timelike Killing vectors could be found. The unique Bunch-Davies vacuum for a comoving observer, matching with Minkowskian vacuum at \emph{arbitrary} short distance, appears as thermal state for a static partner who defines a distinct static vacuum and has no access to the field modes beyond the cosmological horizon \cite{BD}. This information loss with Gibbons-Hawking radiation should certainly cause a decoherence phenomenon for an inertial observer in de Sitter space. Since the thermal spectrum described by a same formula as for the temperature of Hawking radiation and Unruh effect, it is wildly believed \cite{RQI14} that, for RQI in de Sitter space, nothing would change in essentially but only the Hubble parameter $H$ replacing the surface gravity of black hole or the acceleration of Unruh detector. However, as we show in this paper, it is \emph{not} the whole story.

The main point is that the definition of Bunch-Davies vacuum, which relies on an ability to follow a field mode to infinitesimal scales, should be broken near some fundamental scale of quantum gravity. Effectively, this means a boundary condition on the vacuum of comoving observer must be imposed when the momentum of field mode $\vec{k}$ cutoff at Planck scale $\Lambda$. Such constraint can also be interpreted as choosing a harmonic oscillator vacuum at the earliest time $\eta_0(\vec{k})=-\Lambda/H|\vec{k}|$, which eventually results a robust anisotropy signal in CMBR \cite{ALPHA1}. Therefore, in de Sitter space, any quantum information processing with a particular choice of initial vacuum should include those modifications from Planck scale, since it has a directly consequence on the behaviors of quantum correlations \cite{Entropy1,Entropy2,Entropy3,Entropy4,Entropy5,Entropy6}. Even without a complete knowledge about quantum gravity, such investigation on RQI processing may provide a new typical signature to probe Planckian physics  \cite{PLANK}.

Extensive studies show \cite{ALPHA2} that above boundary conditions can be resolved by selecting a non-trivial invariant de Sitter vacuum state called $\alpha$-vacuum, predicted in a free quantum field theory in de Sitter space and can be distinguished by a real number $\alpha$ if the theory consistent with CPT invariance \cite{ALPHA3}. Among this infinite family, a unique element labeled by $\alpha=-\infty$ could be identified with the Bunch-Davies state, over which other vacua $|0^\alpha\ra$ can formally be realized as squeezed state \cite{ALPHA4}, yet each of the vacua $|0^\alpha\ra$ with different UV behavior is the ground state of a different Hilbert space. In the spirit of dS/CFT correspondence \cite{ALPHA5}, these vacua with $\alpha\neq-\infty$ could be related to marginal deformation of boundary CFT. Both bosonic and fermionic $\alpha$-states have been found and play a significant role in cosmology. For instance, fermions in nonthermal state could couple to inflaton field and has significantly larger loop corrections than that would be expected from the Bunch-Davies choice, even when the inflaton itself is in Bunch-Davies vacuum \cite{ALPHA6}. Very recently, a first direct signature of the primordial gravitational wave has been announced by BICEP2 \cite{BICEP1}. To reconcile the BICEP2 result and the PLANCK upper limit, many attempts have been made \cite{BICEP2} by introducing departure from the usual Bunch-Davies case to get enhanced spectra.

In this paper, we analyze the quantum correlations between free quantum field modes (both bosonic and fermionic) and related quantum communication channels in de Sitter space, while the Planckian modification presented by choice of a particular $\alpha$-vacuum. For a static observer, these $\alpha$-vacua become excited and exhibit non-thermal feature derived by the Bogoliubov transformation on his static vacuum. Therefore, in a mostly entangled bipartite system, a deviation from Gibbons-Hawking decoherence should be detected by the inertial observer who is immersed in a bath of non-thermal radiation emanating from the cosmological horizon, while his partner comoving with respect to conformal time see the same entanglement. In this manner, a precise dependence relation between this degradation of quantum correlations and the choice of superselection parameter $\alpha$ would be presented. Comparing with standard Bunch-Davies choice, we find that the possible Planckian physics causes some extra decrement on the quantum correlations. As quantum entanglement can be used to encode the information of these modifications from Planckian physics, it is important to design an operational RQI tasks to probe the physics at fundamental scales \cite{FENG1}. 

On the other hand, many previous RQI works assumed the so-called Single-Mode Approximation (SMA) where the single-particle states of inertial observer are related to states in Rindler basis, therefore the Minkowskian annihilation operator is taken to be one of the right or left moving Unruh modes. Nevertheless, it has been shown \cite{sma} that this does not hold for general states, since the Bogoliubov coefficients appearing in mode-mixing are not localized in frequency. With this in mind, beside the Planckian modification, our investigation on the behavior of quantum correlations is made beyond the SMA, by properly constructing Unruh modes related with $\alpha-$states in global chart of de Sitter space. For instance, for a fixed $\alpha$, we show that bosonic entanglement always vanishes in the infinite-curvature limit with $H\rightarrow\infty$. There is no fundamental difference in the degradation of entanglement for different choices of Unruh mode, which means the entanglement degrades with Hubble scale at same rate. 

In particular, we unveil an unexpected interesting link between choice of $\alpha$ and the asymptotic behavior of fermionic entanglement. For a particular fermionic $\alpha$-vacuum, fermionic entanglement with different choice of Unruh mode would converge to a fixed point $H_c$ determined by $\alpha$. For Bunch-Davies choice with $\alpha\rightarrow-\infty$, we have $H_c\rightarrow+\infty$ consist with the results in Minkowski space \cite{order}, while $H_c$ is finite for non-trivial $\alpha$-vacua. Interestingly, the amount of fermionic entanglement at $H_c$ is a nonvanishing constant independent with choice of Unruh mode and initial vacuum.  

To further explore the physical meaning of the convergent point $H_c$, we employ a particular class of fermionic quantum communication channels, the so-called Grassmann channels \cite{channel1}, for quantum information transmission purposes in de Sitter space. We calculate the classical and quantum capacity which measure the ability of the channels to transmit classical or quantum correlation, and show that the quantum capacity of Grassmann channels would be vanishing at exactly the same critical value $H_c$ where fermionic entanglement converges. Therefore, from the view of quantum information, we conclude that a nontrivial choice of $\alpha$ could be characterized by the vanishing quantum capacity of Grassmann channels in de Sitter spacetime. Nevertheless, as quantum correlation transmission is forbidden beyond $H_c$, we show that the classical capacity of Grassmann channel is always nonnegative unless $H\rightarrow+\infty$, which means classical correlation could be transmitted in the channel until the spacetime curvature approach to infinity. Moreover, we also show that the quantum capacity of a bosonic channel is always nonnegative, that is very different from fermionic case due to their distinct statistics. Nevertheless, by identifying $H_c=\infty$ for bosonic entanglement, we show that the one-to-one correspondence between $H_c$ and the zeros of quantum capacity is universal for both bosonic and fermionic field.

The organization of the paper is as follows. In Sec. II, we first review the Planckian physics in de Sitter space by the boundary condition approach. Then we analyze the thermality of general $\alpha-$vacua by constructing the proper Unruh modes and related Mottola-Allen transformations, based on which the one-particle excited states for bosonic/fermionic field are given explicitly. In Sec. III, we investigate the behavior of quantum correlations of bosonic field in static frame with Planckian modifications and show that the pattern of degradation of quantum correlations is depend on the particular choice of initial vacuum. Similar analysis on the quantum correlations of fermionic fields is given. The convergent feature of entanglement for various $\alpha$ is studied. In Sec. IV, we explore the Grassmann quantum channel for fermionic modes and calculate its classical/quantum capacity. In particular, we find that the convergent point for fermionic entanglement correspond to a quantum channel with vanishing quantum capacity, determined by  a unique Planck-scale cutoff denoted by $\alpha$. Finally, in Sec. V, we summarize our results and discuss some open problems.

\section{Planckian physics in de Sitter space}

\subsection{Boundary condition from Planck scale}
While de Sitter space has the same number of isometries as Minkowski space, quantum field theories in this dynamic background is much complicated by the non-trivial definition of unambiguous vacuum. In the absence of a global timelike Killing vector, many criteria could be applied to choose a particular vacuum. For instance, one could demand that the vacuum state should be annihilated by the generators of the de Sitter isometry group $SO(4,1)$. Here we review the boundary condition approach to Planckian physics in de Sitter space.

Consider a free scalar field in de Sitter space, which satisfy the wave equation $(\Box_g-m^2_{\mbox{\tiny eff}})\phi(x)=0$, where $\Box_g$ is the d'Alembert operator in curved space. The explicit form of the effective mass $m_{\mbox{\tiny eff}}$ depends on its coupling with background. For instance, a conformally coupled scalar field in four-dimension gives $m^2_{\mbox{\tiny eff}}=2H^2$. After the canonical quantization,  the field can be expanded with respect to positive and negative-frequency modes in a Fock representation
\begin{equation}
\phi(x)=\sum_{k}[a_{k}u_{k}(x)+a^{\dag}_{k}u_{k}^{*}(x)]
\label{modeboson}
\end{equation}
The vacuum state is defined by $a_{k}|0\ra=0$ and should respect the spacetime isometries. To specifying mode functions $u_{k}(x)$, one must solve the field equation with some affiliated coordinate systems for different observers.

\begin{figure}[hbtp]
\includegraphics[width=.5\textwidth]{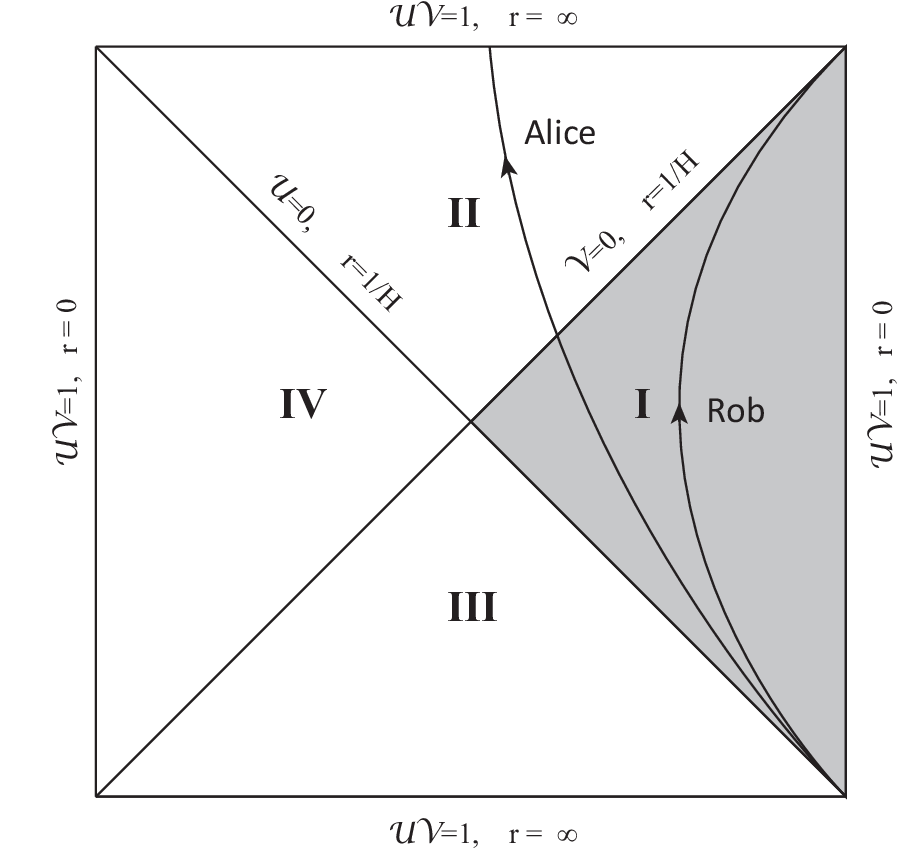}
\caption{The Penrose-Carter diagram of de Sitter space. The static observer situated at $r=0$ would observe the event horizon at $r=H^{-1}$. To construct the proper Unruh modes, the Kruskal coordinates $\mathcal{U}$, $\mathcal{V}$ is also denoted. Considering a bipartite system of Alice and Rob comoving with respect to conformal time initially. If Rob turns to be static in region I while Alice maintains its move with respect to $\eta$, a degradation of quantum correlation due to Gibbons-Hawking effect should be observed for Rob.}
\label{penrose}
\end{figure}

In inflation regime, a conformal observer adopts planar coordinates which reduce the de Sitter metric as
\begin{equation}
ds^2=\frac{1}{(H\eta)^{2}}(-d\eta^{2}+d\rho^{2}+\rho^{2}d\Omega^{2})
\label{confmetric}
\end{equation}
where $\eta=-e^{-Ht}/H$ is conformal time. The coordinates cover the upper right triangle of the Carter-Penrose diagram, while the complete manifold could be covered using the antipodal map $(\eta, \vec{x})\rightarrow(-\eta, \vec{x})$, as depicted in Fig.\ref{penrose}. In inflation regime, one usually considers the coordinate patch with $\eta\in(-\infty,0)$ and traces over the other patch.

Since the space is alway accelerated expanding, it follows that the wavelength of field mode could become arbitrary smaller than the curvature radius if one goes backwards in conformal time far enough.  In the limit $\eta\rightarrow-\infty$, the mode functions reduce to the adiabatic modes defined in flat space and satisfy $\partial_{\eta}u(\eta,\vec{x})=-ik u(\eta,\vec{x})$, which means any distinction between de Sitter space and Minkowski space should be suppressed then. The corresponding Bunch-Davies vacuum $a_{k}(\eta)|0,\eta\ra=0$, which matches the conformal vacuum of Minkowski space in same limit, therefore becomes essentially unique. Since its invariance under de Sitter isometries, the Bunch-Davies vacuum is always chosen as initial state to estimate the primordial power spectrum of inflationary perturbations.

However, as we mentioned before, the existence of fundamental Planck scale, where the unknown quantum gravitational effects become important, prevents us from following a mode back unlimited. To estimate the latest time with Planckian physics \cite{ALPHA1}, a cutoff on physical momentum could be made as $p=k/a(\eta)=\Lambda$, where $\Lambda$ refers to Planckian energy scale . With scale factor $a(\eta)=-1/H\eta$ read from (\ref{confmetric}), one has the conformal time as
\begin{equation}
\eta_{0}=-\frac{\Lambda}{Hk}
\label{bc1}
\end{equation}
when the boundary condition is imposed to redefine the vacuum state for the conformal observer as $a_{k}(\eta_{0})|0,\eta_{0}\ra=0$.

It should be noted that this new vacuum is in general different from Bunch-Davies (but recover the later by taking $\eta_{0}\rightarrow-\infty$) and is a direct result of new physics at Planck scale. Moreover, since the mode expansion (\ref{modeboson}) admits both vacua, this new vacuum could be formally realized as a squeezed state over Bunch-Davies like
\begin{equation}
|0,\eta_{0}\ra=S|0\ra
\label{bc2}
\end{equation}
Without a complete theory of quantum gravity, this boundary condition on vacuum of comoving observer can still provide a typical signature of Planck physics by modify the spectrum of quantum fluctuations after inflation. For instance, it was shown \cite{ALPHA2} that the power spectrum $P(k)\sim\la|u_{k}|^{2}\ra$ with respect to the new vacuum would be modified like $\Delta P(k)/P(k)=\frac{H}{\Lambda}\sin\frac{2\Lambda}{H}$, which is expected to be observed in WMAP or Planck satellite experiments.

More ambitious view is that above argument indeed provides an one-parameter family of vacua with the single parameter given by the fundamental scale (Planckian or stringy). Equivalently, this allows us to discuss the boundary condition from quantum gravity in terms of so-called $\alpha$-vacua which have been known for a long time \cite{ALPHA3,ALPHA6}.

Consider a new set of mode basis related with Bunch-Davies one by Mottola-Allen (MA) transformation
\begin{equation}
u^{\alpha}_{k}(\eta,\vec{x})=N_{\alpha}[u_{k}(\eta,\vec{x})+e^{\alpha}u_{k}^*(\eta,\vec{x})]
\end{equation}
where $\alpha$ is an arbitrary complex number with Re$\alpha<0$, $N_{\alpha}=1/\sqrt{1-e^{\alpha+\alpha^{*}}}$. The one-parameter family of vacuum are defined as $a^{\alpha}_{k}|0^{\alpha}\ra=0$, where
\begin{equation}
a^{\alpha}_{k}=N_{\alpha}[a_{k}-e^{\alpha^{*}}a_{k}^{\dag}]
\label{MAboson}
\end{equation}
are corresponding annihilation operators. These $\alpha$-vacua preserve all $SO(1,4)$ de Sitter isometries. As Re$\alpha\rightarrow-\infty$, $a_{k}^{\alpha}\rightarrow a_{k}$ which indicates the Bunch-Davies vacuum is included in this one-parameter family of vacua. Taking into account that $|0\ra=\bigotimes_k|0_k\ra$ and the commutation relation $[a_k,a_{k'}^\dag]=\delta_{kk'}$, the boundary condition (\ref{bc2}) can now be resolved as a single-mode squeezing transformation
\be
|0^\alpha_k\ra=S(\xi)|0_k\ra=\exp\bigg[\frac{\xi}{2}(a^{\dag}_k a^{\dag}_{k}-a_{k}a_{k})\bigg]|0_k\ra
\label{bcboson}
\ee
where squeezing operator gives the correct MA transformation (\ref{MAboson}) by 
\be
a^{\alpha}_{k}=S(\xi)aS(\xi)^\dag
\label{squez2}
\ee
with squeezing parameter is $\xi=\mbox{arccosh}\;N_\alpha$ . Since the time-reversal invariance of theory requires $\alpha$ be real, we henceforth adopt $\alpha=$Re$\alpha$ for simplicity.

Similarly, a fermionic analogous of $\alpha$-vacua could be defined for a free massive fermionic field, satisfying the wave equation $[i\gamma^\mu(\partial_\mu-\Gamma_\mu)+m_{\mbox{\tiny eff}}]\psi=0$, where $\Gamma_\mu$ is spin connection. The field admits a mode expansion
\begin{equation}
\psi(x)=\sum_{k}[b_{k}u^{+}_{k}(\eta,\vec{x})+c^{\dag}_{k}u^{-}_{k}(\eta,\vec{x})]
\label{conffermion}
\end{equation}
where the creation and annihilation operators for particles and anti-particles obey the anticommutation relations as $\{b_{k},b^{\dag}_{k'}\}=\{c_{k},c^{\dag}_{k'}\}=\delta_{k,k'}$. The Bunch-Davies choice for fermionic modes read as $b_{k}|0_{k}\ra=c_{k}|0_{k}\ra=0$.
With the fermionic MA transformations imposed for both particle and anti-particle operators
\bea
b^{\alpha}_{k}&=&\tilde{N}_{\alpha}[b_{k}-e^{\alpha^{*}}c^{\dag}_{k}]\nonumber\\
c^{\alpha}_{k}&=&\tilde{N}_{\alpha}[c_{k}+e^{\alpha^{*}}b^{\dag}_{k}]
\label{MAfermi}
\eea
where $\tilde{N}_\alpha=1/\sqrt{1+e^{\alpha+\alpha^*}}$. The fermionic $\alpha$-vacua are defined by $b^{\alpha}_{k}|0_{k}^{\alpha}>=c^{\alpha}_{k}|0_{k}^{\alpha}>=0$, which are also invariant under de Sitter isometries. The boundary condition now becomes
\be
|0^\alpha_k\ra=S(\tilde{\xi})|0_k\ra=\exp\bigg[\frac{\tilde{\xi}}{2}(b^{\dag}_k c^{\dag}_{k}+b_{k}c_{k})\bigg]|0_k\ra
\label{bcfermion}
\ee
where squeezing operator gives the correct MA transformation (\ref{MAfermi}) by 
\be
\left(\begin{array}{c}
b^\alpha_k \\
c^{\alpha\dag}_k
\end{array}\right)=S(\tilde{\xi})\left(\begin{array}{c}
b_k \\
c^{\dag}_k
\end{array}\right)S(\tilde{\xi})^\dag
\label{squez3}
\ee
with squeezing parameter is $\tilde{\xi}=\arccos\tilde{N}_\alpha$.

Strictly speaking, only a small set of de Sitter $\alpha$-vacua could be chosen as an alternative initial state rather than Bunch-Davies. The value of $\alpha$ is constrained by the fundamental scale putting in possible quantum gravity model. From (\ref{bc1}), (\ref{bcboson}) and (\ref{bcfermion}), it could be estimated \cite{ALPHA1}
\be
e^\alpha\sim\frac{H}{\Lambda}
\label{estimate}
\ee

\subsection{Thermality and one-particle state from $\alpha$-vacua}

Since the Bunch-Davies vacuum appears thermal with $T=H/2\pi$ for static observer, it is clear that both bosonic and fermionic $\alpha$-vacua would exhibit some non-thermal feature since the corrections from MA transformations (\ref{MAboson}) and (\ref{MAfermi}). Such deviations from pure thermal spectrum are believed to provide a robust signal from Planckian physics in power spectrum of fluctuation after inflation \cite{ALPHA2}. To investigate how these modifications affect RQI processing, we need to obtain the excited states of bosonic/fermionic modes during the particle creation near the cosmological horizon. 

In old RQI literature, a highly idealized scenario so-called the single-mode approximation (SMA) was always employed (see \cite{RQI1} also \cite{RQI7,RQI8}), which means Minkowski annihilation operator is identified to the right Unruh modes, and both the inertial and the non-inertial observer would be sensitive only to a same frequency of quantum fields. However, as was first attained in Ref. \cite{sma}, this may fail to be valid for general states, e.g., a single frequency Minkowski mode is related to a Rindler mode which corresponds to a highly non-monochromatic field excitation.  In parallel with the case of flat space, we relax the SMA in de Sitter space by constructing the Unruh modes for bosonic and fermionic fields respectively.

\subsubsection{Static coordinates and its analytic continuation}

For an inertial observer with static coordinates, de Sitter metric becomes
\be
ds^2=-(1-r^2H^2)dt^2+(1-r^2H^2)^{-1}dr^2+r^2d\Omega^2
\label{static}
\ee
which covers half of the region of conformal coordinates denoted as region I in Fig. \ref{penrose}. The hypersurface on $r=1/H$ is a cosmological causal horizon for an observer situated at $r = 0$. The quantum field modes can be properly separated into positive and negative frequency parts along the trajectory of Killing vector $\partial_{t}$ with respect to cosmological time $t$. 

It is well known that the same quantum field can be expanded with respect to different coordinates. For a static observer situated in region I, we define bosonic mode $u_{\Omega,I}$, fermionic modes $u^+_{\Omega,I}$ and $u^-_{\Omega,I}$, where the subscripts denote the frequency $\Omega=\sqrt{k^2+m^2}$. For completeness, one also need modes in spacelike separated region IV as $u_{\Omega,IV}$, $u^+_{\Omega,IV}$ and $u^-_{\Omega,IV}$. The related field operators should satisfy commutation relation $[a_{\Omega,\Sigma},a_{\Omega',\Sigma'}^\dag]=\delta_{\Omega,\Omega'}\delta_{\Sigma,\Sigma'}$ for bosonic field, and anticommutation relations $\{b_{\Omega,\Sigma},b^{\dag}_{\Omega',\Sigma'}\}=\{c_{\Omega,\Sigma},c^{\dag}_{\Omega',\Sigma'}\}=\delta_{\Omega,\Omega'}\delta_{\Sigma,\Sigma'}$ for fermionic field, where $\Sigma=\{I,IV\}$. Obviously, static vacuum state, which is different from Bunch-Davies one, could be defined by $a_{\Omega,I}|0_{\Omega,I}\ra=0$ and $b_{\Omega,I}|0_{\Omega,I}\ra=c_{\Omega,I}|0_{\Omega,I}\ra=0$. With respect to these modes, the same bosonic field (\ref{modeboson}) can be expanded as
\be
\phi=\sum_{\Omega}[a_{\Omega,I}u_{\Omega,I}+a_{\Omega,I}^\dag u_{\Omega,I}^*+a_{\Omega,IV}u_{\Omega,IV}+a_{\Omega,IV}^\dag u_{\Omega,IV}^*]
\label{b-s}
\ee
and the fermionic field (\ref{conffermion}) has an expansion
\be
\psi=\sum_{\Omega}[b_{\Omega,I}u_{\Omega,I}^++c_{\Omega,I}^\dag u_{\Omega,I}^-+b_{\Omega,IV}u_{\Omega,IV}^++c_{\Omega,IV}^\dag u_{\Omega,IV}^-]
\label{f-s}
\ee

However, it should be noted that above modes are not analytic over whole manifold because the coordinates (\ref{static}) are not geodesically complete. Therefore, it is more convenient to employ Kruskal coordinates $\mathcal{U}$ and $\mathcal{V}$ and make an analytic continuation to the whole de Sitter hyperboloid \cite{GH} 
\be
\mathcal{U}=\mp\exp(-H(t+r_*))\quad,\quad \mathcal{V}=\pm\exp(H(t-r_*))
\ee
where $r_*=\frac{1}{2H}\ln(\frac{1+Hr}{1-Hr})$ is tortoise coordinate, and upper/lower signs refer to right(R)/left(L) Rindler wedges. In these coordinates, the line element (\ref{static}) becomes
\be
ds^2=\frac{1}{H^2(1-\mathcal{U}\mathcal{V})^2}[-4d\mathcal{U}d\mathcal{V}+(1+\mathcal{U}\mathcal{V})^2d\Omega^2]
\label{kruskal}
\ee
which is free of singularity at $\mathcal{U}=\mathcal{V}=0$. While the region I corresponds to the right Rindler wedge R with $\mathcal{U}<0$, $\mathcal{V}>0$, the left Rindler wedge L corresponds to the region IV with $\mathcal{U}>0$, $\mathcal{V}<0$, see Fig. {\ref{penrose}}. 

In rest of this section, we will give all $\alpha-$vacua and their excitation in above $\mathcal{U}-$basis explicitly for both bosonic and fermionic field.

\subsubsection{Bosonic field}

As $e^{i\Omega U}$ is analytic in the lower half $\mathcal{U}-$plane, the Unruh modes in $\mathcal{U}-$basis, which are the combinations of modes with frequency $\Omega$ in Rindler wedges  \cite{UNRUH}
\bea
u_{\Omega,R}&=&\cosh(r_\Omega)u_{\Omega,I}+\sinh(r_\Omega)u_{\Omega,IV}^{*}\no\\
u_{\Omega,L}&=&\cosh(r_\Omega)u_{\Omega,IV}+\sinh(r_\Omega)u_{\Omega,I}^{*}
\label{u-s}
\eea
are also analytic in the lower half $\mathcal{U}-$plane. To derive the correct particle spectrum \cite{exactsolution}, one demands that $\tanh(r_\Omega)=e^{-\pi|k| /H}$, where $|k|$ corresponds the Rindler frequency $\Omega=\sqrt{k^2+m^2}$.

Now the massive scalar field can be expanded with respect to the positive/negative frequency modes in $\mathcal{U}-$basis
\bea
\phi&=&\sum_\Omega(A_{\Omega,R}u_{\Omega,R}+A_{\Omega,R}^{\dag}u_{\Omega,R}^{*}+A_{\Omega,L}u_{\Omega,L}+A_{\Omega,L}^{\dag}u_{\Omega,L}^{*})
\label{unruh}
\eea
Comparing the field expansion in static basis (\ref{b-s}) with the expansion in Unruh basis (\ref{unruh}), we have the Bogoliubov transformation between the Unruh operators $A_{R(L)}$, and static operators $a_{I(IV)}$ 
\bea
A_{\Omega,R}&=&\cosh(r_\Omega)a_{\Omega,I}-\sinh(r_\Omega)a_{\Omega,IV}^{\dag}\no\\
A_{\Omega,L}&=&\cosh(r_\Omega)a_{\Omega,IV}-\sinh(r_\Omega)a_{\Omega,I}^{\dag}
\label{unruh-op}
\eea
and the only nonvanish commutation relation between Unruh operators could be checked likes $[A_{\Omega, R(L)}, A_{\Omega',R(L)}]=\delta_{\Omega, \Omega'}$.

The significant feature of above Unruh modes is that they share the same Bunch-Davies vacuum $|0\ra=|0_U\ra=\bigotimes_{\Omega}|0_{\Omega,U}\ra$ annihilated by $A_{\Omega,R}$ and $A_{\Omega,L}$. To see this, from (\ref{modeboson}) and (\ref{unruh}), we give the transformation between the Unruh modes and Bunch-Davies modes $u_k=\sum_{\Omega}(\alpha^R_{k,\Omega}u_{\Omega,R}+\alpha^L_{k,\Omega}u_{\Omega,L})$, where the coefficients are given by inner product $\alpha^{R(L)}_{k,\Omega}=(u_{\Omega,R(L)},u_k)$. Since the transformation do not mix the creation and annihilation operators \cite{UNRUH, ALPHA4}, i.e., $a_k=\sum_{\Omega}[(\alpha^R_{k,\Omega})^*A_{\Omega,R}+(\alpha^L_{k,\Omega})^*A_{\Omega,L}]$, annihilation operators in $\mathcal{U}-$basis admit the same Bunch-Davies vacuum.
Therefore we could analyze the thermality of Bunch-Davies vacuum by calculating the expectation value of number operator in static frame with respect to Unruh vacuum $|0_U\ra$. For instance, we can show that the Bunch-Davies vacuum is exactly a thermal state in the view of static observer
\bea
\la n\ra_U&\equiv&\la0_{\Omega,U}|a_{\Omega,I}^\dag a_{\Omega,I}|0_{\Omega,U}\ra=\sinh^2(r_\Omega)\no\\
&=&\frac{1}{e^{2\pi|k|/H}-1}=\frac{1}{e^{|k|/T}-1}
\eea
where (\ref{unruh-op}) is employed and Gibbons-Hawking temperature is defined by $T=H/2\pi$.

To generalize above analysis on thermality to $\alpha-$vacua (\ref{bcboson}), similar as in (\ref{MAboson}), we impose MA transformation on Unruh operators $A_{R(L)}$
\bea
A^\alpha_{\Omega,R}=N_\alpha(A_{\Omega,R}-e^\alpha A_{\Omega,L}^\dag)\no\\
A^\alpha_{\Omega,L}=N_\alpha(A_{\Omega,L}-e^\alpha A_{\Omega,R}^\dag)
\label{unruh-op-a}
\eea
which should share the common $\alpha-$vacua as $|0^\alpha\ra\equiv|0_{U}^\alpha\ra=\bigotimes_\Omega|0_{\Omega,U}^\alpha\ra$. Its thermality could be checked with respect to these $\alpha-$vacua in $\mathcal{U}-$basis by
\bea
\la \bar{n}\ra_U&\equiv&\la0_{\Omega,U}^\alpha|a_{\Omega,I}^\dag a_{\Omega,I}|0_{\Omega,U}^\alpha\ra=\frac{(\sinh r_\Omega+e^\alpha \cosh r_\Omega)^2}{1-e^{2\alpha}}\no\\
&=&\frac{1}{e^{|k|/T}-1}\frac{(1+e^{\alpha+|k|/2T})^2}{1-e^{2\alpha}}
\label{bthermal}
\eea
representing a non-thermal spectrum as we expected. From (\ref{unruh-op}), (\ref{unruh-op-a}) and the definition $A^\alpha_{\Omega,R}|0_{\Omega,U}^\alpha\ra=0$, after a bit of algebra, the explicit form of the $\alpha-$vacua in $\mathcal{U}-$basis is
\bea
|0_{\Omega,U}^\alpha\ra&=&[N_\alpha(\cosh r_\Omega+e^\alpha\sinh r_\Omega)]^{-1}\times\exp\bigg[\frac{\tanh r_\Omega +e^\alpha}{1+e^\alpha\tanh r_\Omega}a^\dag_{\Omega,I}a^\dag_{\Omega,IV}\bigg]|0_{\Omega,I},0_{\Omega,IV}\ra\no\\
&=&\sqrt{1-\tanh^2(r_\Omega)\Delta^2}\sum_{n=0}^\infty \tanh^nr_\Omega\Delta^n|n_{\Omega,I},n_{\Omega,IV}\ra
\label{unruh-alpha}
\eea
where $\Delta\equiv\frac{1+e^\alpha\tanh^{-1}r_\Omega}{1+e^\alpha\tanh  r_\Omega}=\frac{1+e^{\alpha+\pi |k|/H}}{1+e^{\alpha-\pi|k|/H}}$ 
agreeing with the results in \cite{ALPHA4}. As $\alpha\rightarrow-\infty$, these corrections can be neglected and standard Bunch-Davies vacuum is recovered.

To obtain the one-particle excitation of above vacua, one need apply the most general creation operator of $\alpha-$modes which can be written as a linear combination
\be
A^{\alpha\;\dag}_{\Omega,U}=q_L A_{\Omega,L}^{\alpha\;\dag}+q_R A_{\Omega,R}^{\alpha\;\dag}
\ee
where two real parameters (for simplicity) have $q_L^2+q_R^2=1$. From (\ref{unruh-op}), (\ref{unruh-op-a}) and (\ref{unruh-alpha}), we have
\bea
|1^\alpha_{\Omega,U}\ra&\equiv& A^{\alpha\;\dag}_{\Omega,U}|0^\alpha_{\Omega,U}\ra\no\\
&=&(1-\tanh^2r_\Omega\Delta^2)\sum_{n=0}^\infty \tanh^nr_\Omega\Delta^n\sqrt{n+1}(q_L|n_{\Omega,I},n_{\Omega,IV}+1\ra
+q_R|n_{\Omega,I}+1,n_{\Omega,IV}\ra)\no\\
\label{unruh-one}
\eea
which has a symmetric form on two Rindler wedges. If we restrict our view to $q_L=0$ sector, the results in SMA regime could be recovered.

In summary, in order to investigate the thermality of general $\alpha$-vacua defined for conformal observer, we have constructed the Unruh modes (\ref{u-s}) analytic over whole spacetime manifold. The related Unruh operators (\ref{unruh-op}) admit the same Bunch-Davies vacuum which appears as a pure thermal state in the view of static observer. After MA transformation, the new Unruh operators  (\ref{unruh-op-a}) share the same $\alpha$-vacua which become a two-mode state (\ref{unruh-alpha}) with non-thermal spectrum for static observer. The most general one-particle excitation over $\alpha$-vacua is given in (\ref{unruh-one}), along with (\ref{unruh-alpha}) would be employed in Sec. \ref{correlations} to investigate the behavior of bosonic quantum correlations in de Sitter space.

\subsubsection{Fermionic field}

In the context of fermionic field, similar Unruh modes as (\ref{u-s}) could be constructed by analytic continuation to entire de Sitter manifold in Kruskal coordinates. The corresponding annihilation operators are \cite{sma}
\bea
B_{\Omega,R}=\cos(\tilde{r}_\Omega)b_{\Omega,I}-\sin(\tilde{r}_\Omega)c_{\Omega,IV}^\dag&,&
B_{\Omega,L}=\cos(\tilde{r}_\Omega)b_{\Omega,IV}-\sin(\tilde{r}_\Omega)c_{\Omega,I}^{\dag}\no\\
C_{\Omega,R}=\cos(\tilde{r}_\Omega)c_{\Omega,IV}+\sin(\tilde{r}_\Omega)b_{\Omega,I}^\dag&,&
C_{\Omega,L}=\cos(\tilde{r}_\Omega)c_{\Omega,I}+\sin(\tilde{r}_\Omega)b_{\Omega,IV}^\dag
\label{u-s-f}
\eea
where $\tan(\tilde{r}_\Omega)=e^{-\pi|k|/H}$ and $|k|$ corresponds the Rindler frequency $\Omega=\sqrt{k^2+m^2}$. From the anticommutation relations of fermionic operators in static chart, one can deduce that $\{B_{\Omega,\sigma},B_{\Omega',\sigma'}^\dag\}=\delta_{\Omega,\Omega'}\delta_{\sigma,\sigma'}$, $\{C_{\Omega,\sigma},C_{\Omega',\sigma'}^\dag\}=\delta_{\Omega,\Omega'}\delta_{\sigma,\sigma'}$ ($\sigma=\{R,L\}$) with all other anticommutators vanish.

Same as bosonic case, the Bunch-Davies operators are proportional to Unruh operators like $b_k=\alpha(B_{\Omega,R}\otimes \mathbf{1}_R)+\beta^*(\mathbf{1}_R\otimes B_{\Omega,L})$ (similar relation for antiparticle operator), therefore share the same Bunch-Davies vacuum \cite{ALPHA4}. If the fermionic Fock basis is decomposed as $|0\ra=|0_U\ra=\bigotimes_{\Omega}|0_{\Omega,R}\ra\otimes|0_{\Omega,L}\ra$, one has $B_{\Omega,R}|0_{\Omega,R}\ra=C_{\Omega,R}|0_{\Omega,R}\ra=0$ and $B_{\Omega,L}|0_{\Omega,L}\ra=C_{\Omega,L}|0_{\Omega,L}\ra=0$. From (\ref{u-s-f}), we can show that the fermionic Bunch-Davies vacuum also appears thermal in the view of a static observer. For instance, the detector of a static observer in region I would register a particle spectrum like
\bea
\la n\ra_U&\equiv&\la0_{\Omega,R}|b^\dag_{\Omega,I} b_{\Omega,I}|0_{\Omega,R}\ra=\sin^2(\tilde{r}_\Omega)\no\\
&=&\frac{1}{1+e^{2\pi|k|/H}}=\frac{1}{1+e^{|k|/T}}
\label{f-bd}
\eea
where (\ref{u-s-f}) is used.

For general fermionic $\alpha-$vacua, we construct the following Unruh operators by MA transformation
\bea
B^\alpha_{\Omega,R}=\tilde{N}_\alpha(B_{\Omega,R}-e^\alpha C^\dag_{\Omega,R})&,&
B^\alpha_{\Omega,L}=\tilde{N}_\alpha(B_{\Omega,L}-e^\alpha C^\dag_{\Omega,L})\no\\
C^\alpha_{\Omega,R}=\tilde{N}_\alpha(C_{\Omega,R}+e^\alpha B^\dag_{\Omega,R})&,&
C^\alpha_{\Omega,L}=\tilde{N}_\alpha(C_{\Omega,L}+e^\alpha B^\dag_{\Omega,L})
\label{fermialpha}
\eea
which share the same $\alpha-$vacua  $|0^\alpha\ra\equiv|0_{U}^\alpha\ra=\bigotimes_\Omega|0^\alpha_{\Omega,R}\ra\otimes|0^\alpha_{\Omega,L}\ra$, annihilated by Unruh operators $B^\alpha_{\Omega,R}|0^\alpha_{\Omega,R}\ra=C^\alpha_{\Omega,R}|0^\alpha_{\Omega,R}\ra=0$ and $B^\alpha_{\Omega,L}|0^\alpha_{\Omega,L}\ra=C^\alpha_{\Omega,L}|0^\alpha_{\Omega,L}\ra=0$. 
The non-thermal feature of fermionic $\alpha-$vacua can be deduced similar as in (\ref{f-bd}), e.g. from (\ref{u-s-f}), (\ref{fermialpha}), the particle spectrum registered by a static observer in region I is
\bea
\la \bar{n}\ra_U&\equiv&\la0^\alpha_{\Omega,R}|b^\dag_{\Omega,I} b_{\Omega,I}|0^\alpha_{\Omega,R}\ra=\frac{(\sin\tilde{r}_\Omega+e^\alpha\cos\tilde{r}_\Omega)^2}{1+e^{2\alpha}}\no\\
&=&\frac{1}{e^{|k|/T}+1}\frac{(1+e^{\alpha+|k|/2T})^2}{1+e^{2\alpha}}
\label{fthermal}
\eea

Comparing with bosonic case (\ref{unruh-alpha}), the fermionic $\alpha-$vacua could also be expressed as two-mode squeezed state but with more subtle 
\bea
|0^\alpha_{\Omega,R}\ra&=&\sum_n F_{n}|n_{\Omega,I}\ra^+|n_{\Omega,IV}\ra^-\no\\
|0^\alpha_{\Omega,L}\ra&=&\sum_n G_{n}|n_{\Omega,I}\ra^-|n_{\Omega,IV}\ra^+
\label{r-l-vac}
\eea
The simplest case preserving the fermionic characteristics is Grassmann scalar, which reduce the occupation number to one, i.e. $F_n=G_n=0$ ($n\geqslant2$). Since these vacua are annihilated by operators (\ref{fermialpha}), the coefficients could be determined
\bea
F_0&=&G_0=\tilde{N}_\alpha(\cos\tilde{r}_\Omega-e^\alpha \sin\tilde{r}_\Omega)\no\\
F_1&=&-G_1=\tilde{N}_\alpha(\sin\tilde{r}_\Omega+e^\alpha \cos\tilde{r}_\Omega)
\label{coe}
\eea
The explicit form of the fermionic $\alpha-$vacua is
\bea
|0_{\Omega,U}^\alpha\ra&=&|0^\alpha_{\Omega,R}\ra\otimes|0^\alpha_{\Omega,L}\ra\no\\
&=&\tilde{N}_\alpha^2[(\cos\tilde{r}_\Omega-e^\alpha \sin\tilde{r}_\Omega)^2|0000\ra_\Omega-(\sin\tilde{r}_\Omega+e^\alpha \cos\tilde{r}_\Omega)^2|1111\ra_\Omega\no\\
&+&(e^\alpha\cos(2\tilde{r}_\Omega)+\sin(2\tilde{r}_\Omega)(1-e^{2\alpha})/2)\times(|0011\ra_\Omega+|1100\ra_\Omega)]
\label{unruh-f-0}
\eea 
where we introduce the notations
\be
|1111\ra_\Omega=b^\dag_{I}c^\dag_{IV}c^\dag_{I}b^\dag_{IV}|0_{\Omega,I}\ra^+|0_{\Omega,IV}\ra^-|0_{\Omega,I}\ra^-|0_{\Omega,IV}\ra^+
\label{notation}
\ee

The 1st-excitated states of particle sector are obtained by applying general particle creation operator which is a linear combination 
\be
B^{\alpha\;\dag}_{\Omega,U}=q_R (B^{\alpha\;\dag}_{\Omega,R}\otimes \mathbf{1}_R)+q_L (B^{\alpha\;\dag}_{\Omega,L}\otimes \mathbf{1}_L)
\ee
where $q_R^2+q_L^2=1$. With in mind the definition (\ref{r-l-vac}) and (\ref{notation}), we have
\bea
|1^\alpha_{\Omega,R}\ra&=&B^{\alpha\;\dag}_{\Omega,R}|0^\alpha_{\Omega,R}\ra=|1_{\Omega,I}\ra^+|0_{\Omega,IV}\ra^-\no\\
|1^\alpha_{\Omega,L}\ra&=&B^{\alpha\;\dag}_{\Omega,L}|0^\alpha_{\Omega,L}\ra=|0_{\Omega,I}\ra^-|1_{\Omega,IV}\ra^+
\eea
Therefore, the one-particle states are
\bea
|1_{\Omega,U}^\alpha\ra&=&q_R |1^\alpha_{\Omega,R}\ra\otimes|0^\alpha_{\Omega,R}\ra  +q_L |0^\alpha_{\Omega,R}\ra\otimes|1^\alpha_{\Omega,R}\ra  \no\\
&=&q_R\tilde{N}_\alpha[(\cos\tilde{r}_\Omega-e^\alpha \sin\tilde{r}_\Omega)|1000\ra_\Omega-(\sin\tilde{r}_\Omega+e^\alpha \cos\tilde{r}_\Omega)|1011\ra_\Omega]\no\\
&+&q_L\tilde{N}_\alpha[(\sin\tilde{r}_\Omega+e^\alpha \cos\tilde{r}_\Omega)|1101\ra_\Omega+(\cos\tilde{r}_\Omega-e^\alpha \sin\tilde{r}_\Omega)|0001\ra_\Omega]
\label{unruh-f-one}
\eea

It should be noted \cite{order} that different operator ordering in fermionic systems could lead to nonunique results when computing entanglement measures for the same state. For instance, if we rearrange operator ordering in (\ref{notation}) as $b^\dag_{I}c^\dag_{I}c^\dag_{IV}b^\dag_{IV}$, then a new Fock basis is defined $|1111\ra_\Omega'=-|1111\ra_\Omega$, which usually results a different amount of quantum entanglement. This so-called physical ordering, in which all region I operators appear to the left of all region IV operators, was proposed in \cite{order2} to guarantee the entanglement behavior of above states would yield physical results.

In summary, we construct fermionic Unruh operators (\ref{u-s-f}) which admit same Bunch-Davies vacuum of a free fermionic field in de Sitter space, and generalize them to the form (\ref{fermialpha}) corresponding to $\alpha$-vacua. Both operators exhibit correct thermal/non-thermal feature as shown in (\ref{f-bd}) and (\ref{fthermal}), therefore give explicitly expression of fermionic $\alpha$-vacua (\ref{unruh-f-0}) and its one-particle excitation (\ref{unruh-f-one}).

\subsubsection{Single-mode approximation}

As we mentioned before, the SMA used in old RQI literature is not correct for general states. Nevertheless, the choice of Unruh modes indeed makes no essential differences in computation of certain quantum correlation measure (e.g. quantum discord \cite{discord2}). For completeness, we list the $\alpha-$vacua and their one-particle states in SMA regime in follow, which agree with previous work \cite{FENG3}.

For bosonic field, SMA leads to $a_k \sim A_{\Omega,R}$, or $q_L=0$ equivalently. The bosonic $\alpha-$vacua (\ref{unruh-alpha}) and their 1st excitation (\ref{unruh-one}) become
\bea
|0_{\Omega}^\alpha\ra&=&\sqrt{1-\tanh^2r_\Omega\Delta^2}\sum_{n=0}^\infty \tanh^nr_\Omega\Delta^n|n_{\Omega,I},n_{\Omega,IV}\ra\label{sma-b-0}\\
|1^\alpha_{\Omega}\ra&=&(1-\tanh^2r_\Omega\Delta^2)\sum_{n=0}^\infty \tanh^nr_\Omega\Delta^n\sqrt{n+1}|n_{\Omega,I}+1,n_{\Omega,IV}\ra
\label{sma-b-1}
\eea

For fermionic field, similar restriction could be made by $q_L=0$. The vacuum states of particle sector are
\bea
|0^\alpha_{\Omega}\ra&=&\tilde{N}_\alpha(\cos\tilde{r}_\Omega-e^\alpha \sin\tilde{r}_\Omega)|0_{\Omega,I}\ra^+|0_{\Omega,IV}\ra^-+\tilde{N}_\alpha(\sin\tilde{r}_\Omega+e^\alpha \cos\tilde{r}_\Omega)|1_{\Omega,I}\ra^+|1_{\Omega,IV}\ra^-\no\\
&=&\sum_{n=0}^{1}\frac{\tan^n\tilde{r}_\Omega\tilde{\Delta}^n}{\sqrt{1+\tan^2\tilde{r}_\Omega\tilde{\Delta}^2}}|n_{\Omega,I}\ra^+|n_{\Omega,IV}\ra^-
\label{sma-f-0}
\eea
where $\tilde{\Delta}\equiv\frac{1+e^\alpha\tan^{-1}  \tilde{r}_\Omega}{1-e^\alpha\tan  \tilde{r}_\Omega}=\frac{1+e^{\alpha+\pi |k|/H}}{1-e^{\alpha-\pi |k|/H}}$. The one-particle states are
\be
|1^\alpha_{\Omega}\ra=|1_{\Omega,I}\ra^+|0_{\Omega,IV}\ra^-
\label{sma-f-1}
\ee

\section{quantum entanglement in de Sitter space}
\label{correlations}

In above analysis, we have illustrated that the $\alpha$-vacua, which resolve the boundary condition (\ref{bc2}) from physics at Planck scale, do exhibit the nonthermal feature for both bosonic and fermionic field as shown in (\ref{bthermal}) and (\ref{fthermal}). Since the field modes in region IV beyond cosmological event horizon are unaccessible for static observer in region I, the related degree of freedom should be traced over. The main point is such information-loss should also suffer these nonthermal corrections from possible Planckian physics denoted by $\alpha$, along with the choice of different Unruh modes by real parameter $q_L$, which make the behavior the quantum correlations among field modes be very different from our expectation. 

\subsection{Entanglement negativity}
\label{3.1}

To estimate the quantum correlation, we should carefully choose a proper entanglement measure. While it is easy for pure state, the state of affairs is much involved in case of mixed states (see \cite{measure1} for a review). Consider a bipartite state $\rho$, it is called entangled if it cannot be written as $\rho=\sum_{i}p_i\rho_i^A\otimes\rho_i^B$, a tensor product of reduced density matrices with probability distribution $\sum_{i}p_i=1$. This physically means the state cannot be produced using only local operation and classical communication (LOCC). The intricate nature of entanglement for mixed states lies in several respects. For instance, unlike pure entangled states that can be completely characterized by the violation of Bell inequalities, there exists mixed entangled state can nevertheless be described by a local hidden variable model \cite{measure2}. On the other hand, the entanglement entropy of reduced density matrices widely used to measure entanglement of pure state would involve both classical and quantum correlation for a mixed state, therefore is no longer a good entanglement measure \cite{measure4}. 

A key property of any quantity to be a good measure of entanglement is so-called entanglement monotone, which means the quantity should not increase under LOCC \cite{measure1}. Many such kind of measures have been proposed for mixed states, however most of them cannot be computed since they are given by variational expressions over possible LOCC protocols. In this section, we consider a computable entanglement measure called entanglement negativity \cite{nega}, defined from the partial transpose criterion \cite{measure3} that provides a sufficient criterion for entanglement. Given a general bipartite density matrix in the tensor product Hilbert space $\mathcal{H}_A\otimes\mathcal{H}_B$ as $\rho_{ik,jl}=\la a_ib_k|\rho|a_jb_l\ra$, where $|a_i\ra$ and $|b_k\ra$ with $i\in\{1,2,\cdots,\mbox{dim}(\mathcal{H}_A)\}$ and $k\in\{1,2,\cdots,\mbox{dim}(\mathcal{H}_B)\}$ are basis of each subsystem, one defines the partial transpose operation with respect to the one of the systems, e.g., subsystem $A$
\be
\rho^\Gamma_{ik,jl}\equiv\rho_{il,jk}=\la a_ib_l|\rho|a_jb_k\ra
\label{PPT}
\ee
If $\rho^\Gamma$ has at least one eigenvalue of the partial transpose is negative, then the density matrix is entangled. However a state with positive partial transpose (PPT) can still be entangled, called bond entanglement or undistillable entanglement which means it is not possible to extract pure entanglement from it using only LOCC. Therefore, one could define negativity as measure of distillable entanglement based on the amount of violation of the PPT criterion as
\be
N(\rho)=\frac{||\rho^\Gamma||_1-1}{2}
\ee
where the trace-norm of an operator is $||\mathcal{O}||_1\equiv\mbox{Tr}(\sqrt{\mathcal{O}^\dag\mathcal{O}})$. From the definition, the negativity provides a measure of the number of negative eigenvalues $\lambda_i$ of the density matrix the sum of negative eigenvalues, hence gives \cite{nega}
\be
N(\rho)=\frac{1}{2}\sum_i(|\lambda_i|-\lambda_i)=-\sum_{\lambda_i<0}\lambda_i
\label{negat}
\ee
It has been proved \cite{measure5} that above quantity is indeed do not increase under any LOCC, i.e., an entanglement monotone, therefore could be an appropriate measure of entanglement. Moreover, since negativity could be related to the maximal fidelity achieved in a teleportation protocol that uses $\rho$ as a resource \cite{nega,measure6}, it also has an important practical meaning in quantum information process.

In the relativistic frame, negativity and its logarithmic form for a quantum system undergoes a relativistic motion are observer-dependent \cite{RQI7,RQI8,RQI10}. This can be illustrated by a bipartite system sharing any maximally entangled Bell states initially, e.g., $|\Phi\ra=(|0\ra_A|0\ra_B+|1\ra_A|1\ra_B)/\sqrt{2}$, which from (\ref{negat}) gives negativity $N(|\Phi\ra)=0.5$. Once one of the subsystem undergoes an uniform acceleration $a$, the related field modes should be transformed through Bogoliubov transformation. The tracing over field modes beyond acceleration horizon in general gives a new mixed bipartite state. Depending on the acceleration parameter ($r_b=\mbox{arccosh}(1-e^{-|k|/T_U})^{-1/2}$ for bosonic field and $r_f=\mbox{arccos}(1+e^{-|k|/T_U})^{-1/2}$, where Unruh temperature is $T_U=a/2\pi$), negativity would be degraded with increasing Unruh temperature. While negativity $N(|\Phi\ra)=0.5$ for both bosonic and fermionic fields without acceleration, they have different asymptotic negativity when $T_U\rightarrow\infty$, i.e., $N_{\tiny\mbox{bosonic}}\rightarrow0$ when $r_b\rightarrow\infty$ (which means in this limit the total correlations consist of classical correlations plus bound entanglement), but residual fermionic negativity $N_{\tiny\mbox{fermionic}}\rightarrow0.25$ when $r_f\rightarrow\frac{\pi}{4}$.

In the rest of this section, we investigate both bosonic and fermionic entanglement in de Sitter space using negativity as a proper entanglement measure. In particular, beyond SMA, we demonstrate that fermionic negativity converges with same residual entanglement $N_{\tiny\mbox{fermionic}}=0.25$ but at different $r_\Omega$ depending on the choice of different $\alpha-$vacua. Furthermore, in order to consist with the analysis on quantum communication in Sec. \ref{channel}, at the end of the section, we explore the operational meaning of this fermionic residual entanglement by quantify the maximally possible violation of the Clauser- Horne-Shimony-Holt (CHSH) inequality \cite{chsh}.

\subsection{Bosonic entanglement}

We first consider the scalar modes with infinite levels, from which a qubit can be truncated \cite{truncation} \footnote{To have a well-defined states like (\ref{tripartite}) and effective quantum field theory in de Sitter space, the modes with the largest values of $k$ must be chosen leads to a $4$th order adiabatic state \cite{eft}.}.
The simplest way to illustrate the influence of $\alpha$-vacua on quantum information task is to investigate a maximally entangled state between different frequency
\be
|\Phi\ra=\frac{1}{\sqrt2}(|0^{\alpha}_s\ra\otimes|0^\alpha_{\Omega,U}\ra+|1^{\alpha}_s\ra\otimes|1^\alpha_{\Omega,U}\ra)
\label{tripartite}
\ee
shared by two conformal observers Alice and Rob with respect to conformal time $\eta$ initially. We assume that Alice has a detector only sensitive to mode $s$, while Rob has detector sensitive to $\Omega$ corresponding to an Unruh state. As Rob turning to be static while Alice maintain her comoving trajectory with respect to $\eta$, each Unruh states should be transformed to static coordinates like (\ref{unruh-alpha}) and (\ref{unruh-one}). The density matrix of whole tripartite state $\rho_{A,I,IV}=|\Phi\ra\la\Phi|$ includes modes on both sides of event horizon along with modes in planar coordinates. Since Rob has no access to field modes beyond event horizon, we should trace over all states in region IV which results a mixed state. From (\ref{unruh-alpha}) and (\ref{unruh-one}), we have
\bea
\rho_{A,I}&=&\mbox{Tr}_{IV}\rho_{A,I,IV}\no\\
&=&\sum_{n=0}^{\infty}\frac{1}{2}\tanh^{2n}r_\Omega\Delta^{2n}(1-\tanh^2r_\Omega\Delta^2) \Big[\;|0n\rangle \langle 0n|+\sqrt{(n+1)(1-\tanh^2r_\Omega\Delta^2)}(q_R|1n+1\rangle\langle 0n|\no\\
&+&q_L\tanh r_\Omega\Delta|1n\ra\langle0n+1|)+\sqrt{(n+1)(n+2)}(1-\tanh^2r_\Omega\Delta^2)q_Rq_L|1n+2\ra\la 1n|\no\\
&+ &(n+1 )(1-\tanh^2r_\Omega\Delta^2)(q_R^2|1n+1\rangle \langle
1n+1|+q_L^2|1n\rangle \langle
1n|)+H.C.\;\Big]
\label{domAI}
\eea
where we use the notation $|nm\ra\equiv|n^\alpha_s\ra\otimes|m_{\Omega,I}\ra$.

To estimate the quantum correlation, we should calculate the negativity (\ref{negat}) which is an entanglement monotone defined as the sum of negative eigenvalues $\lambda_i$ of partial transposed density matrix. From (\ref{PPT}) and (\ref{domAI}), we have its partial transposed form
\bea
\rho^\Gamma_{A,I}&=&\sum_{n=0}^{\infty}\frac{1}{2}\tanh^{2n}r_\Omega\Delta^{2n}(1-\tanh^2r_\Omega\Delta^2) \Big[\;|0n\rangle \langle 0n|+\sqrt{(n+1)(1-\tanh^2r_\Omega\Delta^2)}(q_R|1n\rangle\langle 0n+1|\no\\
&+&q_L\tanh r_\Omega\Delta|1n+1\ra\langle0n|)+\sqrt{(n+1)(n+2)}(1-\tanh^2r_\Omega\Delta^2)q_Rq_L|1n\ra\la 1n+2|\no\\
&+ &(n+1 )(1-\tanh^2r_\Omega\Delta^2)(q_R^2|1n+1\rangle \langle
1n+1|+q_L^2|1n\rangle \langle
1n|)+H.C.\;\Big]
\label{ptdomAI}
\eea

For $q_R=1$, the transposed matrix has a block-diagonal structure and the eigenvalues could be given analytically. It follows that the negativity with SMA regime is
\bea
N_{A,I}&=&\sum_{n=0}^{+\infty}\frac{1}{4}\tanh^{2n}r_\Omega\Delta^{2n}(1-\tanh^2r_\Omega\Delta^2)\no\\
&\times&\Bigg|\sqrt{\bigg[\tanh^2r_\Omega\Delta^2+n(\frac{\coth^2r_\Omega}{\Delta^2}-1)\bigg]^2+4(1-\tanh^2r_\Omega\Delta^2)}
-\tanh^2r_\Omega\Delta^2+n\Big(\frac{1-\coth^2r_\Omega}{\Delta^2}\Big)\Bigg|\no\\
\label{bnegativity}
\eea
which is a function of $\alpha$ and Hubble scale.

\begin{figure}[hbtp]
\includegraphics[width=.7\textwidth]{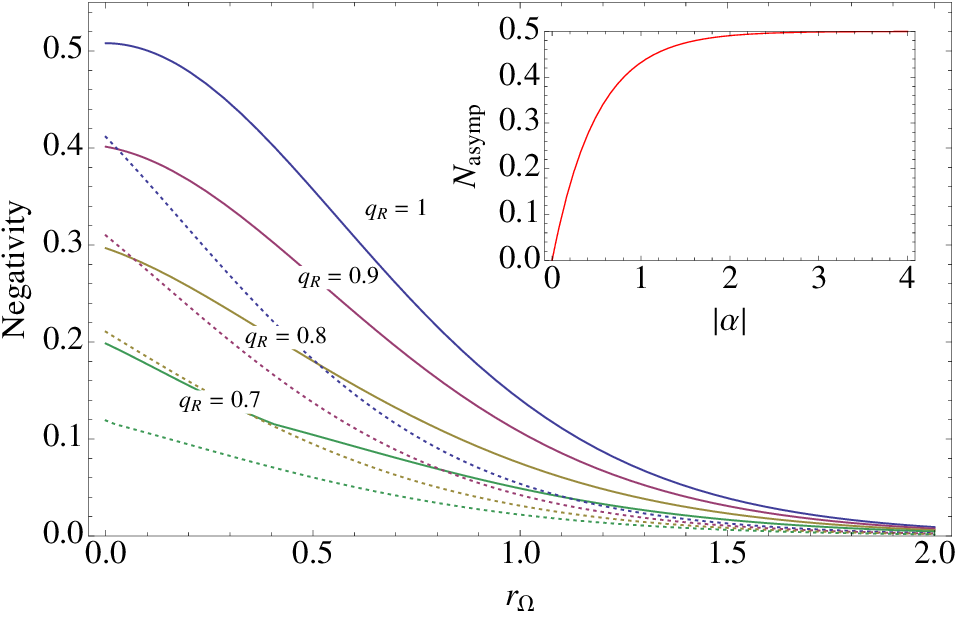}
\caption{The negativity of Alice-I system as the function of Hubble scale for various choices of $q_R$ and $\alpha$. The parameter $r_\Omega$ is defined by $
\tanh(r_\Omega)=e^{-\pi|k| /H}
$ where $|k|$ corresponds the Rindler frequency $\Omega=\sqrt{k^2+m^2}$ and $H$ is time-independent Hubble scale. The continuous solid (dashed) curves from top to bottom correspond to $\alpha=-6$ ($\alpha=-3$) and $q_R=1,0.9,0.8,0.7$, respectively. In particular, the blue solid and dashed curves represent the degradation of entanglement with SMA assumed. The asymptotic negativity for $r_\Omega\rightarrow0$ is also depicted, where the residual bosonic negativity is $N_{\mbox{\tiny asymp}}=0$ for $\alpha\rightarrow 0^-$.}
\label{negativity1}
\end{figure}

For general $q_R\neq1$, the non-diagonal terms appearing in (\ref{ptdomAI}) make us fail to solve the eigenvalues analytically, but can estimate them numerically. To compare the SMA case (\ref{bnegativity}), we depict the negativity between Alice and Rob with different choice both of $q_R$ and $\alpha$ as in Fig.\ref{negativity1}.

Our first observation is that a degradation of quantum entanglement occurs for static Rob (see Fig.\ref{negativity1}) as we expected. This phenomenon roots from the information-loss via de Sitter radiation detected by inertial observer similar as Unruh effect for accelerated frame in flat space. In particular, there is no essential difference in the pattern of entanglement behavior for different choice of Unruh modes with fixed $\alpha$. Therefore, we would focus on the influence of choice of $\alpha-$states. For simplicity, we choose $q_L=0$ which recovers the SMA regime.

For Bunch-Davies state which means $\alpha=-\infty$, the spectrum is pure thermal. Recall that \cite{exactsolution} $
\tanh(r_\Omega)=e^{-\pi|k| /H}
$ where $|k|$ corresponds the Rindler frequency $\Omega=\sqrt{k^2+m^2}$. If $H\rightarrow0$ ($r_\Omega\rightarrow0$), de Sitter space approaches flat with infinite large curvature radius $l\equiv1/H\rightarrow\infty$, and the negativity would approach 0.5 due to vanish radiation then. On the other hand, in the limit of infinite curvature $H\rightarrow\infty$ ($r_\Omega\rightarrow\infty$), the state has no distillable entanglement since negativity is exactly zero.

The new feature of entanglement in de Sitter space is that the Planckian modification represented by vacua ambiguity could be directly encoded in quantum correlations measured by static observer Rob. Comparing with the standard Bunch-Davies choice, the negativity is suppressed for all $\alpha$-vacua choice with $\alpha\neq-\infty$. In a realistic model \cite{ALPHA2} with $H\sim10^{14}$GeV and $\Lambda\sim10^{16}$GeV, we can estimate the typical value of $\alpha$ from (\ref{estimate}) as $\alpha\sim-3.5$. This choice of initial vacuum results a significant modification in the degradation of entanglement for Rob. Since the quantum entanglement encodes the possible Planckian physics, we expect that, in principle, one can probe the unknown physics at Planck scale by some quantum information tasks \cite{FENG1}.

It should be remarked that, unlike in Bunch-Davies choice, for a non-trivial $\alpha$-vacuum, even in the limit $r_\Omega\rightarrow0$ where de Sitter space approaching flat, the negativity is still smaller than 0.5 which means the distillable entanglement suffers a decrement for static observer. At first glance, this seems contradict with RQI in flat space \cite{RQI7} where an inertial observer should maintain his amounts of entanglement. However, since $\alpha$-vacua become squeezed states over Bunch-Davies vacuum which matches the conformal vacuum of flat space in Minkowskian limit, state (\ref{tripartite}) with $\alpha\neq-\infty$ now appears mixed for inertial observer Rob. To demonstrate above intuition, here we sketch how to estimate the asymptotic negativity of general $\alpha$-vacua as $r_\Omega\rightarrow0$ directly from the interpretation of $\alpha$-vacua as single-mode squeezed states built over the Bunch-Davies state \cite{squez}. Given $r_\Omega\rightarrow0$, the spacetime approaches flat which implies that the Bunch-Davies vacuum matches the conformal vacuum of Minkowski space, which is now the proper vacuum state for the static observer Rob. By the squeezing operator $S(\alpha)=\exp[(a^{\dag}_k a^{\dag}_{k}-a_{k}a_{k})\xi/2]$ in (\ref{bcboson}), for a static observer, a general $\alpha$-vacuum should be transformed in the perspective of Rob as \cite{squez,METHOD}
\bea
|0^\alpha_k\ra&=&\exp\bigg[\frac{1}{2}a^{\dag2}\tanh\xi\bigg]\exp\bigg[-\frac{1}{2}(aa^{\dag}+a^\dag a)\ln(\cosh\xi)\bigg]|0_{M,k}\ra\no\\
&=&(1-e^{2\alpha})^{1/4}\sum_{n=0}^\infty\frac{\sqrt{(2n)!}}{2^nn!}e^{n\alpha}|2n_{M,k}\ra
\label{asymp1}
\eea
and one-particle excitation is
\bea
|1^\alpha_k\ra&=&a^{\alpha\dag}_k|0^\alpha_k\ra=(1-e^{2\alpha})^{3/4}\sum_{n=0}^\infty\frac{\sqrt{(2n)!}}{2^nn!}e^{n\alpha}\sqrt{2n+1}|(2n+1)_{M,k}\ra
\label{asymp2}
\eea
where subscript $M$ denotes Minkowski limit and the squeezing parameter is $\xi=\mbox{arccosh}\;N_\alpha$ as in (\ref{squez2}). Substitute (\ref{asymp1}) and (\ref{asymp2}) into the maximally entangled state (\ref{tripartite}), the transformed density matrix is  
\bea
\rho_{AR}&=&\frac{1}{2}\rho_{nm}(\alpha)\bigg[\sqrt{1-e^{2\alpha}}\big(\sqrt{2m+1}|0,2n\ra\la1,2m+1|+\sqrt{2n+1}|1,2n+1\ra\la0,2m|\big)\no\\
&&+|0,2n\ra\la0,2m|+(1-e^{2\alpha})|1,2n+1\ra\la1,2m+1|\bigg]\no\\
\rho_{nm}(\alpha)&=&\sqrt{1-e^{2\alpha}}\sum_{n,m}e^{(n+m)\alpha}\sqrt{(2n)!(2m)!}/2^{n+m}n!m!
\eea
which is an infinite dimensional matrix with notation $|nm\ra\equiv|n^\alpha\ra|m_M\ra$. After partial transposing, the negativity could be numerically estimated by sum all its negative eigenvalues, which clearly only depends on the value of $\alpha$. In particular, for a $\alpha$-vacuum with $\alpha\rightarrow-\infty$, corresponding vanishing squeezing parameter, (\ref{asymp1}) gives $|0^\alpha_k\ra=|0_{M,k}\ra$, therefore no degradation on negativity should be expected, i.e., $N=0.5$ from (\ref{tripartite}). Otherwise, the nontrivial $\alpha$-vacua induce a decrement for static observer. For $\alpha\rightarrow0^-$, the bosonic negativity is $N_{\mbox{\tiny asymp}}=0$ (see Fig. \ref{negativity1}). This analysis shows that our result is indeed consistent with those arguments of RQI in flat space.

Finally, we would like to compare our result with other entanglement measures. As mentioned in Sec. \ref{3.1}, entanglement entropy can mix both classical and quantum correlation for a mixed state (e.g. in an high temperature mixed state, it gives an extensive result for the thermal entropy that has nothing to do with entanglement), thus is not a good entanglement measure from the perspective of quantum information. This is also evident from the fact that the entanglement entropy of subsystem $S_A$ is no longer equal to $S_R$ for mixed bipartite system. Nevertheless, a symmetric mutual information could be constructed by $I_{A:R}=S_A+S_R-S_{A\cup R}$ which measures the total classical and quantum correlations in bipartite system. From (\ref{domAI}), the reduced density matrix for Alice and Bob could be given by $\rho_A=\mbox{Tr}_{I}(\rho_{A,I})$ and $\rho_I=\mbox{Tr}_{A}(\rho_{A,I})$. By the definition $S_{A(I)}=-\mbox{Tr}(\rho_{A(I)}\ln\rho_{A(I)})$, the mutual information of $\rho_{AI}$ could be easily obtained, which is \cite{FENG3}
\bea
I_{A:I}&=&S(\rho_A)+S(\rho_I)-S(\rho_{A,I})\no\\
&=&1-\frac{1}{2}\log_{2}\left( \tanh ^{2}r_\Omega\Delta^2\right) -\frac{1}{2}(1-\tanh^2r_\Omega\Delta^2)\no\\
&&\sum_{n=0}^{\infty}\tanh ^{2n}r_\Omega\Delta^{2n}\Big\{(1-n+n\frac{\coth^2r_\Omega}{\Delta^2})\log_2(1-n+n\frac{\coth^2r_\Omega}{\Delta^2})\no\\
&&-[n+2-(n+1)\tanh^2r_\Omega\Delta^2]\log_2[n+2-(n+1)\tanh^2r_\Omega\Delta^2]\Big\}
\eea
Recall that the parameter $r_\Omega=\mbox{arctanh}e^{-\pi|k| /H} $ and $\Omega=\sqrt{k^2+m^2}$, mutual information is dependent on the choice of $\alpha$, Hubble scale $H$ and mass of field. Nevertheless, one should note that although above calculation of mutual information may share similar properties as in Ref. \cite{Entropy3,Entropy6}, we emphasize more on the quantum informational methodology in curved spacetime \cite{RQI6,RQI7} rather than holographic techniques used in quantum gravity. Moreover, since (\ref{tripartite}) is pure, one reads that $S(\rho_{A,I})=S(\rho_{IV})$ and $S(\rho_{A,IV})=S(\rho_{I})$. Therefore the combined mutual information, $I_{A,I}+I_{A,IV}=2$, is conserved which suggests a correlation transfer between Alice-RobI and Alice-RobIV systems \cite{RQI7} and be independent with initial vacua selection.

\subsection{Fermionic entanglement}

Since all matter fields in nature are fermionic, it is important to extend above results to quantum correlations between fermionic field modes. For this purpose, we assume two comoving observers Alice and Rob with respect to conformal time $\eta$ share a maximally entangled initial state as
\be
|\Psi\ra=\frac{1}{\sqrt2}(|0^{\alpha}_s\ra\otimes|0_{\Omega,U}^\alpha\ra+|1^{\alpha}_s\ra\otimes|1_{\Omega,U}^\alpha\ra)
\label{fermitripartite}
\ee
Similar as bosonic case, we assume that Alice has a detector only sensitive to mode $s$, while Rob has detector sensitive to $\Omega$ corresponding to an Unruh state. While Rob turning to be static, the field modes beyond cosmological horizon become causal disconnected for him. Therefore, the initial pure state $\rho=|\Psi\ra\la\Psi|$ appear as a mixed state for Rob by tracing over the degrees of region IV. To estimate the amount of fermionic quantum entanglement, we need to calculate the negativity (\ref{negat}) for fermionic state (Grassmann scalar for simplicity). 

From Unruh states (\ref{unruh-f-0}) and (\ref{unruh-f-one}), we choose the so-called physical ordering of fermionic operators \cite{order}. The density matrix of Alice-Rob system becomes
\bea
\rho_{A,I}&=&\frac{1}{2}\Big[c^{4}|000\ra\la000|+s^2c^2(|010\ra\la010|+|001\ra\la001|)+s^4|011\ra\la011|\no\\
&+&q_R^2(c^2|110\ra\la110|+s^2|111\ra\la111|)+q_L^2(s^2|110\ra\la110|+c^2|100\ra\la100|)\no\\
&+&q_R(c^3|000\ra\la110|+s^2c|001\ra\la111|)-q_L(c^2s|001\ra\la100|-s^3|011\ra\la110|)\no\\
&-&q_R q_L sc|111\ra\la100|+H.C.\Big]
\label{dmfermi}
\eea
with the shorthand notation $|nmp\ra\equiv|n^\alpha_s\ra\otimes|m_{\Omega,I}\ra^+\otimes|p_{\Omega,I}\ra^-$ and
\be
c=\tilde{N}_\alpha(\cos \tilde{r}_\Omega- e^\alpha\sin \tilde{r}_\Omega)\qquad,\qquad
s=\tilde{N}_\alpha(\sin \tilde{r}_\Omega+e^\alpha\cos \tilde{r}_\Omega)
\ee
We have seen that such fermionic system is much simpler than the bosonic case due to its statistics. Similar as before, after partial transposing (\ref{dmfermi}), we have
\bea
\rho_{A,I}^\Gamma&=&\frac{1}{2}\Big[c^{4}|000\ra\la000|+s^2c^2(|010\ra\la010|+|001\ra\la001|)+s^4|011\ra\la011|\no\\
&+&q_R^2(c^2|110\ra\la110|+s^2|111\ra\la111|)+q_L^2(s^2|110\ra\la110|+c^2|100\ra\la100|)\no\\
&+&q_R(c^3|010\ra\la100|+s^2c|011\ra\la101|)-q_L(c^2s|000\ra\la101|-s^3|010\ra\la111|)\no\\
&-&q_R q_L sc|100\ra\la111|+H.C.\Big]
\label{dmfermipt}
\eea

To quantify fermionic entanglement, we use the negativity which sums over all negative eigenvalues of (\ref{dmfermipt}). To demonstrate the pattern of entanglement behavior, we study its dependence on the choice of Unruh modes and $\alpha-$states, respectively.

\begin{figure}[hbtp]
\includegraphics[width=.7\textwidth]{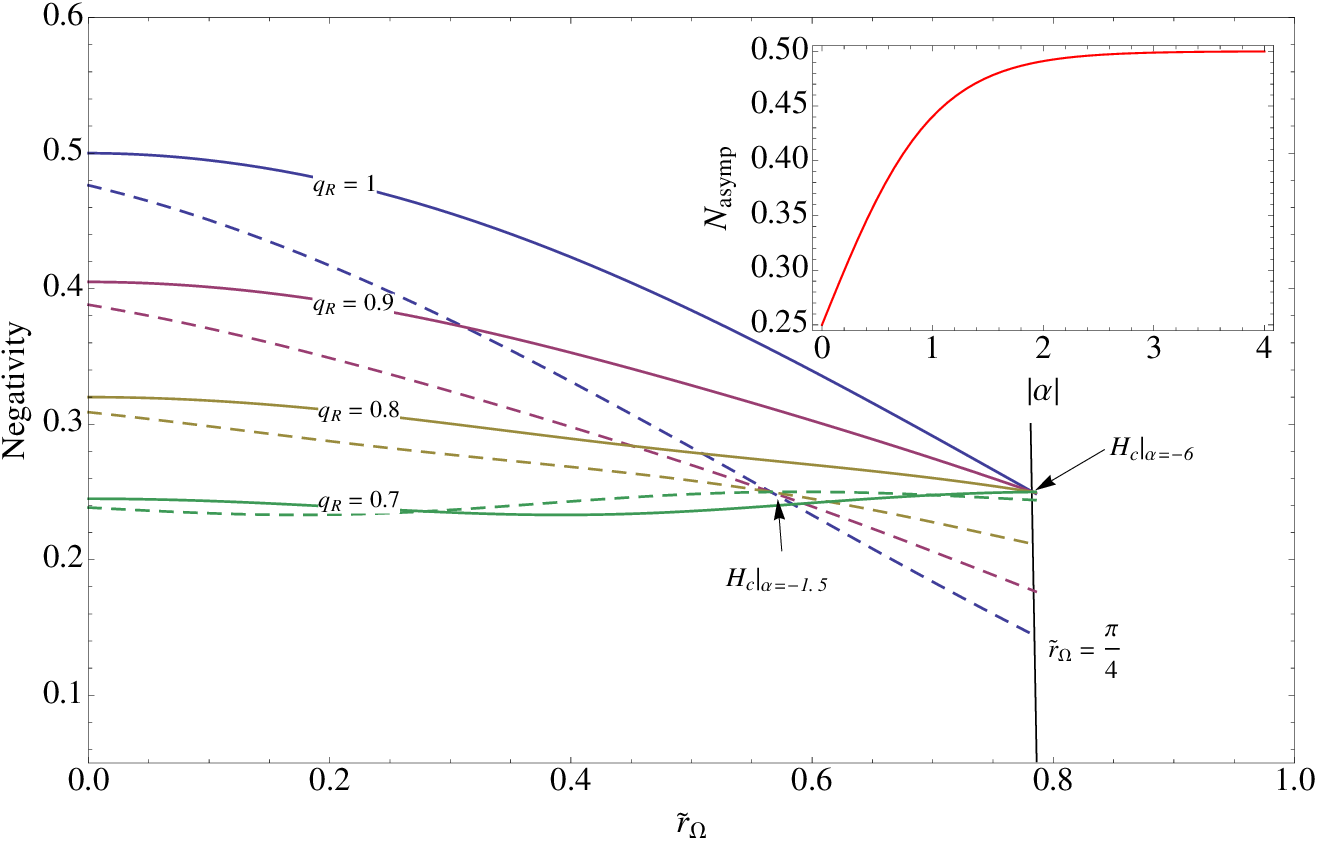}
\caption{The negativity of Alice-I system as the function of Hubble scale for various choices of $q_R$ and $\alpha$. The parameter $\tilde{r}_\Omega$ is defined by $
\tan(\tilde{r}_\Omega)=e^{-\pi|k| /H}
$ where $|k|$ corresponds the Rindler frequency $\Omega=\sqrt{k^2+m^2}$ and $H$ is time-independent Hubble scale. The continuous solid (dashed) curves from top to bottom correspond to $\alpha=-6$ ($\alpha=-1.5$) and $q_R=1,0.9,0.8,0.7$, respectively. In particular, the blue solid and dashed curves represent the degradation of entanglement with SMA assumed. The asymptotic negativity for $\tilde{r}_\Omega\rightarrow0$ is also depicted, where residual negativity $N_{asymp}=0.25$ for $\alpha\rightarrow0^-$.}
\label{ferminegativity}
\end{figure}

We first consider the simple case in which SMA with $q_R=1$ is assumed. Only one negative eigenvalue exist for the partial transposed matrix, which gives the negativity
\be
N_{A,I}=\frac{1}{2}(\tan^2\tilde{r}_\Omega\tilde{\Delta}^2+1)^{-1}
\ee
For $\alpha\rightarrow-\infty$, Rob recovers the Bunch-Davies choice. Recall that \cite{exactsolution} $\tan(\tilde{r}_\Omega)=e^{-\pi|k| /H}$, in Minkowskian limit $H=0$ ($\tilde{r}_\Omega=0$) where space approaching flat, the state (\ref{fermitripartite}) remains maximally entangled. As the space curvature grows infinitely $H\rightarrow\infty$ ($\tilde{r}_\Omega\rightarrow\frac{\pi}{4}$), using Fermi statistics $\tan\tilde{r}=\exp(-\pi|k|/H)$, the negativity is $\frac{1}{4}$ which means the state (\ref{fermitripartite}) will always preserve some degree of distillable entanglement. For $\alpha\neq-\infty$, the quantum entanglement is more suppressed compared with Bunch-Davies case. As depicted in Fig. \ref{ferminegativity}, the negativity is smaller than 0.5 even for Minkowskian limit since the $\alpha$-vacua then become squeezed states over Minkowskian vacuum. Similar as bosonic case, this asymptotic negativity can also be directly deduced from the interpretation of $\alpha$-vacua as single-mode squeezed states built over the Bunch-Davies state. For $\tilde{r}_\Omega\rightarrow0$, one only need to consider the squeezing operation (\ref{squez3}) with $S(\tilde{\xi})=\exp[(b^{\dag}_k c^{\dag}_{k}+b_{k}c_{k})\tilde{\xi}/2]$ and squeezing parameter $\tilde{\xi}=\arccos\tilde{N}_\alpha$. The $\alpha$-states become \cite{METHOD}
\be
|0^\alpha_{k}\ra=\frac{1}{\sqrt{1+e^{2\alpha}}}(|0_k\ra^+|0_k\ra^-+e^{\alpha}|1_k\ra^+|1_k\ra^-)\qquad,\qquad
|1^\alpha_{k}\ra=|1_{k}\ra^+|0_{k}\ra^-
\ee
where superscript $\pm$ indicates particle and anti-particle. Substitute above equation into (\ref{fermitripartite}) and restrict to particle sector, we have density matrix
\be
\rho_{AR}=\frac{1}{2+2e^{2\alpha}}[|00\ra\la00|+|11\ra\la11|+e^{2\alpha}|01\ra\la01|+\sqrt{1+e^{2\alpha}}(|00\ra\la11|+|11\ra\la00|)]
\ee
which after partial transposing can give the asymptotic negativity as
\be
N_{\mbox{\tiny asymp}}(\alpha)=\frac{1}{2+2e^{2\alpha}}
\ee
which reads $0.5$ for Bunch-Davies choice $\alpha\rightarrow-\infty$, but degrades as $\alpha\rightarrow0^-$ and unlike bosonic case with residual fermionic negativity $0.25$ (see Fig. \ref{ferminegativity}).

On the other hand, as $H\rightarrow\infty$, less amount of distillable entanglement than that in Bunch-Davies case could be preserved for Rob. In an extreme case with $\alpha\rightarrow0^-$, the residual entanglement at limit $H\rightarrow\infty$ can even become vanish which seems be very different with the fermionic RQI in flat space \cite{FENG3}. Similar as before, we can estimate an alternative entanglement measure using entanglement entropy, i.e., fermionic mutual information which measures total classical and quantum correlations. Straightforwardly, the mutual information of the state (\ref{dmfermi}) is given by
\bea
I_{A,I}&=&1+\frac{1}{2}\tilde{N}^2_\alpha(\cos \tilde{r}_\Omega- e^\alpha\sin \tilde{r}_\Omega)^2\big[(\tan^2\tilde{r}_\Omega\tilde{\Delta}^2+2)\log_{2}(\tan^2\tilde{r}_\Omega\tilde{\Delta}^2+2)\no\\
&-&(2\tan^2\tilde{r}_\Omega\tilde{\Delta}^2+1)\log_{2}(2\tan^2\tilde{r}_\Omega\tilde{\Delta}^2+1)+\tan^2\tilde{r}_\Omega\tilde{\Delta}^2\log_{2}\tan^2\tilde{r}_\Omega\tilde{\Delta}^2\big]
\eea
Similar as bosonic case, since $S(\rho_{A,I})=S(\rho_{IV})$ and $S(\rho_{A,IV})=S(\rho_{I})$, therefore the combined mutual information, $I_{A,I}+I_{A,IV}=2$, is conserved quantity which suggests a correlation transfer between Alice-RobI and Alice-RobIV systems should be independent with initial vacua selection.

However, as we mentioned before, the complete understanding on the behavior of fermionic entanglement should be built beyond SMA. For Unruh modes with varying $q_R$, we observe the convergent behavior of fermionic negativity, depicted in Fig. \ref{ferminegativity}. Denoting the convergent point as $H_c$ and the corresponding negativity as $N_0$, we observe that $H_c\rightarrow\infty$ for Bunch-Davies states with $N_0=0.25$, similar as what happens for RQI in Minkowski space \cite{order}. Moreover, for a general $\alpha-$state ($\alpha\neq-\infty$), the entanglement would converge at a finite $H_c$, beyond which the negativity becomes divergent again. Since the amount of entanglement at $H_c$ is always $N_0=0.25$, independent with the choice of $\alpha$, we find that the convergent point is constrained by
\be
H_c=\pi|k|\bigg(\ln\frac{1+e^\alpha}{1-e^\alpha}\bigg)^{-1}
\label{critical}
\ee
which gives $\tilde{r}_{\Omega}=0.56$ if $\alpha=-1.5$ and $\tilde{r}_{\Omega}=0.43$ if $\alpha=-1$ for instance. For the fermionic mode with fixed frequency, relation (\ref{critical}) implies that the more deviation of initial fermionic vacuum from Bunch-Davies choice the earlier the fermionic negativity converges. This definitely is a new feature for quantum correlation in de Sitter space when fundamental minimal length is considered. Moreover, in Sec. \ref{channel}, we would reexam this entanglement convergence from the view of quantum communication channel, and show that the convergent points (For bosonic entanglement, $H_c\rightarrow\infty$ is independent with $\alpha$, while $H_c$ should satisfy (\ref{critical}) for fermionic entanglement) have an interesting link to the ability of quantum channel to transmit quantum correlations.

\subsection{Residue entanglement and nonlocality}

Before we move to the issue of quantum communication, we would like to first explore the operational nature of the nonzero fermionic residual entanglement at $H_c$, by applying the criterion introduced in Ref. \cite{crit} to quantify the maximally possible violation of the Clauser-Horne-Shimony-Holt (CHSH) inequality \cite{chsh}. 

Apparently, the residual fermionic entanglement cannot be attributed to bound entanglement, since the negativity measures distillable entanglement. Therefore, it is natural to ask whether there are some quantum information tasks using the quantum correlations remaining at $H_c$ can outperform states with appropriate classical correlations. Otherwise, we think the entanglement beyond $H_c$ is not interesting, since it has no operational meaning in de Sitter space. To demonstrate this question, we should check that if the residual fermionic entanglement can be used to violate the CHSH inequality, which is the optimal Bell inequality for this situation.

\begin{figure}[hbtp]
\includegraphics[width=.7\textwidth]{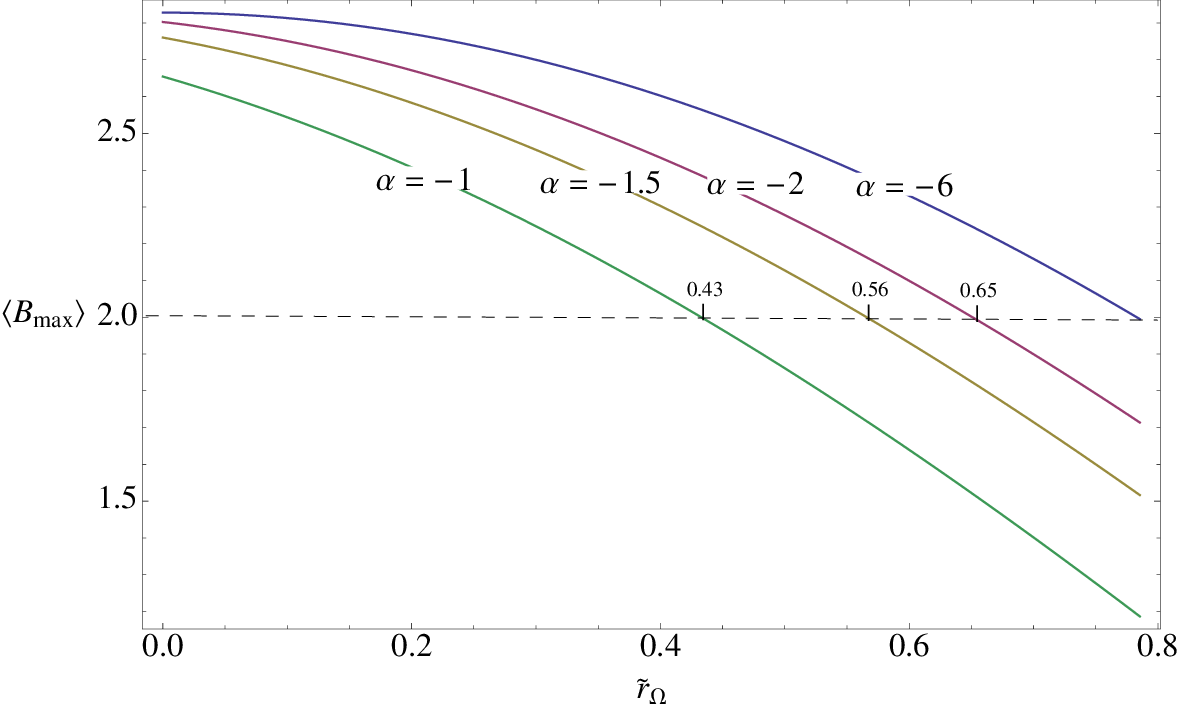}
\caption{The maximally possible violation of CHSH inequality $\la\mathcal{B}_{\mbox{\tiny max}}\ra_\rho$ as a function of Hubble scale. The curves from top to bottom correspond to $\alpha=-6,-2,-1.5,-1$, respectively. Beyond the convergent points $H_c$, the surviving entanglement cannot be used to violate the CHSH inequality, therefore no quantum information tasks using these correlations can outperform states with appropriate classical correlations.  }
\label{chshb}
\end{figure}

Starting from (\ref{dmfermi}) and restrict further to particle sector w.o.l.g., we have two-qubit state
\bea
\rho_{A,I}^+&=&\frac{1}{2}\bigg[c^2|00\ra\la00|+s^2|01\ra\la01|+(q_R^2+q_L^2s^2)|11\ra\la11|+q_L^2c^2|10\ra\la10|\no\\
&&+q_Rc(|00\ra\la11|+|11\ra\la00|)\bigg]
\label{chshrho}
\eea
where we use the shorthand notation $|nm\ra\equiv|n_s^\alpha\ra\otimes|m_{\Omega,I}\ra^+$ and superscript $+$ means the restriction on particle sector. In an experiment to test local realistic theories, Bell inequalities or its generalization the CHSH inequality provide a bound like \cite{chsh}
\be
|\la\mathcal{B}_{\mbox{\tiny CHSH}}\ra_\rho|\leqslant2
\ee
where $\mathcal{B}_{\mbox{\tiny CHSH}}=\mathbf{ a}\cdot\sigma\otimes(\mathbf{ b}+\mathbf{b }')\cdot\sigma+\mathbf{ a}'\cdot\sigma\otimes(\mathbf{ b}-\mathbf{b }')\cdot\sigma$, $\mathbf{ a}$, $\mathbf{ a}'$, $\mathbf{ b}$, $\mathbf{ b}'$ are unit vectors in $\mathbb{R}^3$, and $\sigma$ is the vector Pauli matrices. For some choices of unit vectors, the inequality can be violated by quantum states up to the value $2\sqrt2$. It was shown \cite{crit} that the maximally possible value $\la\mathcal{B}_{\mbox{\tiny max}}\ra_\rho$ of the CHSH expectation value for a given two-qubit state $\rho$ is determined by
\be
\la\mathcal{B}_{\mbox{\tiny max}}\ra_\rho=2\sqrt{\mu_1+\mu_2}
\ee
where $\mu_1$, $\mu_2$ are the two largest eigenvalues of $U(\rho)=T^T_\rho T_\rho$. The correlation matrix $T=(t_{ij})$ of the generalized Bloch decomposition of the density operator $\rho$ is given by $t_{ij}=\mbox{Tr}[\rho\sigma_i\otimes\sigma_j]$. For the density matrix (\ref{chshrho}) we find
\be
U(\rho_{A,I}^+)=\frac{q_R^2(\cos \tilde{r}_\Omega- e^\alpha\sin \tilde{r}_\Omega)^2}{1+e^{2\alpha}}\left(\begin{array}{ccc}1 & 0 & 0 \\0 & 1 & 0 \\0 & 0 & \frac{(\cos \tilde{r}_\Omega- e^\alpha\sin \tilde{r}_\Omega)^2}{1+e^{2\alpha}}
\end{array}\right)
\ee
Thus we obtain
\be
\la\mathcal{B}_{\mbox{\tiny max}}\ra_\rho=2\sqrt{2}\frac{q_R(\cos \tilde{r}_\Omega- e^\alpha\sin \tilde{r}_\Omega)}{\sqrt{1+e^{2\alpha}}}
\ee
Since we only concern about possible violation of inequality by the residual entanglement at the convergent points $H_c$, we depict the maximally possible value $\la\mathcal{B}_{\mbox{\tiny max}}\ra_\rho$ in Fig. \ref{chshb} under SMA with $q_R=1$.

We find that beyond the convergent points $H_c$, the residual fermionic entanglement cannot violate CHSH inequality, which means that no quantum information tasks using these correlations can outperform states with appropriate classical correlations. In an operational perspective, this implies \cite{crit2} that the comoving observer Alice in de Sitter space cannot communicate on a quantum information level with a static observer Rob resting with respect to the cosmological horizon. With this understanding, we conclude that it is more important to investigate quantum correlation in de Sitter space in a operational framework, rather than simply quantifying the amount of entanglement, since not all kind of entanglement can be used as a resource in real quantum information process. In next section, we would convince this intuition by constructing quantum communication channels in de Sitter space and show that channels' quantum capacity is zero at convergent points $H_c$.

\section{Quantum communication in de Sitter space}
\label{channel}

In a certain sense, above analysis directly on entanglement measure is more or less naive, since what one really concerns is how to utilize quantum entanglement as resource in a quantum information process, e.g., quantum teleportation \cite{RQI7,FENG1}. Therefore, we now turn to study a more physical scenario by constructing quantum communication channels between Alice and Rob in de Sitter space, and investigate the influence from the different choice of $\alpha$. Such analysis would demonstrate the physical meaning of convergent point $H_c$ more clearly. 

We estimate two of the most investigated quantities, i.e., the classical and quantum capacity of a quantum channel, which measure the ability of a quantum channel to transmit classical or quantum correlations. In particular, we consider the sender Alice (comoving with respect to conformal time $\eta$) who optimally prepares some states encoded in terms of Unruh modes, and the receiver Rob (inertial motion with respect to cosmic time $t$) able to intercept and decode the transmitted information with arbitrarily high precision but in terms of Rindler modes.

Following the notation of \cite{channel1}, let $\mathcal{B}(\mathcal{H})$ denote the algebra of linear operators acting on a $d$-dimensional Hilbert space $\mathcal{H}\equiv \mathbb{C}^d$, a quantum channel is a completely positive trace-preserving map $\mathcal{K}:\mathcal{B}(A')\rightarrow\mathcal{B}(A) $, mapping density operators from an input Hilbert space $\mathcal{H}_{A'}$ to an output Hilbert space $\mathcal{H}_{A}$. To circumvent the problem that the field theories prescribed by Alice and Rob are not unitarily equivalent \cite{WALD}, we work in isometric picture which is constituted by the isometric extension of channel $V_\mathcal{K}:A'\mapsto AC$ into a higher-dimensional Hilbert space and transformation through unitary operator $U_{\mathcal{K}}:A'C'\mapsto AC$.

For a quantum channel $\mathcal{K}$, the capacity could be given by the regularized expression \cite{CAPACITY1}. For instance, the classical capacity is 
\be
C(\mathcal{K})=\lim_{n\rightarrow\infty}\frac{1}{n}C_{\mbox{\tiny Hol}}(\mathcal{K}^{\otimes n})
\label{ccap}
\ee
where $C_{\mbox{\tiny Hol}}(\mathcal{K})$ is the Holevo quantity. The quantum capacity is 
\be
Q(\mathcal{K})=\lim_{n\rightarrow\infty}\frac{1}{n}Q^{(1)}(\mathcal{K}^{\otimes n})
\label{qcap}
\ee
where $Q^{(1)}(\mathcal{K})$ is the optimized coherent information.
 
\subsection{Unruh channel}

As a warm up, we consider an Unruh channel of bosonic field which is investigated in {\cite{channel2}} for Minkowski space. Since RQI in de Sitter space with Bunch-Davies choice has no essential difference with Minkowskian RQI \cite{RQI14}, results from later case could be directly applied for the same channel in de Sitter space with $\alpha=-\infty$. For instance, the unitary operator $U_{AC}$ assigning a two-mode entangled Rindler state to every Unruh state is
\be
U_{AC}(r_\Omega)=\exp[r_\Omega(a^\dag c^\dag-ac)]
\label{uop}
\ee
where a different mode notation $A\equiv I$ and $C\equiv IV$ are adopted, and operator $a$ satisfies canonical communication relation $[a_k,a_k^\dag]=\delta(k-k')$ (similar relation hold for $c$). The unitary transformation do produces the correct state in the view of static observer, which is
\be
U_{AC}(r_\Omega)|n\ra_A|0\ra_C=\frac{1}{\cosh^{n+1}r_\Omega}\sum_{m=0}^\infty\left(\begin{array}{c}n+m \\n\end{array}\right)^\frac{1}{2}\tanh^mr_\Omega|n+m\ra_A|m\ra_C
\ee

For general $\alpha$, MA-transformations (\ref{unruh-op-a}) introduce additional parameter $\alpha$ and can be realized as a squeezing transformation. This enables us to effectively introduce the unitary operator 
\be
U^\alpha_{AC}(r_\Omega)=\exp\bigg[\mbox{arctanh}\bigg(\frac{\tanh r_\Omega+e^\alpha}{1+e^\alpha\tanh r_\Omega}\bigg)(a^\dag c^\dag-ac)\bigg]
\label{uopa}
\ee
which has a distinct parameter from that in \cite{channel2} and becomes (\ref{uop}) as $\alpha\rightarrow-\infty$. By using identity \cite{METHOD}
\bea
U^\alpha_{AC}(r_\Omega)&=&\big(1-\tanh^2r_\Omega\Delta^2\big)^\frac{1}{2}\exp\bigg[\frac{\tanh r_\Omega+e^\alpha}{1+e^\alpha\tanh r_\Omega}a^\dag c^\dag\bigg]\no\\
&&\times\exp\bigg[\frac{1}{2}\ln(1-\tanh^2r_\Omega\Delta^2)(a^\dag a+ c^\dag c)\bigg]\exp\bigg[-\frac{\tanh r_\Omega+e^\alpha}{1+e^\alpha\tanh r_\Omega}a  c \bigg]
\label{id}
\eea
we have the action of bosonic unitary operator
\be
U^\alpha_{AC}(r_\Omega)|n\ra_A|0\ra_C=(1-\tanh^2r_\Omega\Delta^2)^\frac{n+1}{2}\sum_{m=0}^\infty\left(\begin{array}{c}n+m \\n\end{array}\right)^\frac{1}{2}\bigg(\frac{\tanh r_\Omega+e^\alpha}{1+e^\alpha\tanh r_\Omega}\bigg)^m|n+m\ra_A|m\ra_C
\ee 
which is consistent with (\ref{unruh-alpha}).

To transform a bosonic qudit with multi-rail encoding
\be
|\Psi\ra_{A'C'}=|\Psi\ra_{A'}|0\ra_{C'}=\sum_{i=1}^d\beta_ia_i^\dag|0\ra_{A'}|0\ra_{C'}=\sum_{i=1}^d\beta_i|i\ra_{A'}|0\ra_{C'}\ee
$d$ copies of unitary operator should be used. After some straightforward algebra, the final output state is
\bea
|\Phi\ra_{AC}&=&\bigotimes_{i=1}^dU^\alpha_{A_iC_i}|\Psi\ra_{A'C'}\no\\
&=&(1-\tanh^2r_\Omega\Delta^2)^\frac{d+1}{2}\sum_{k=1}^\infty\bigg(\frac{\tanh r_\Omega+e^\alpha}{1+e^\alpha\tanh r_\Omega}\bigg)^{k-1}\sum_I\bigg[\sum_{i=1}^d\beta_i\sqrt{l_{i}+1}|I^{(i)}\ra_A|I\ra_C\bigg]
\eea 
where $|I\ra_C=|l_1,\cdots,l_d\ra_C$ is a multi-index labeling fro basis states of the completely symmetric subspace of $(k-1)$ photons in $d$ modes, and $|I^{(i)}\ra_A=|l_{1},\cdots,l_{i}+1,\cdots,l_{d}\ra_A$ denotes that $k$ photons in $d$ modes contained in subsystem $A$.

\begin{figure}[hbtp]
\centering
\includegraphics[width=.7\textwidth]{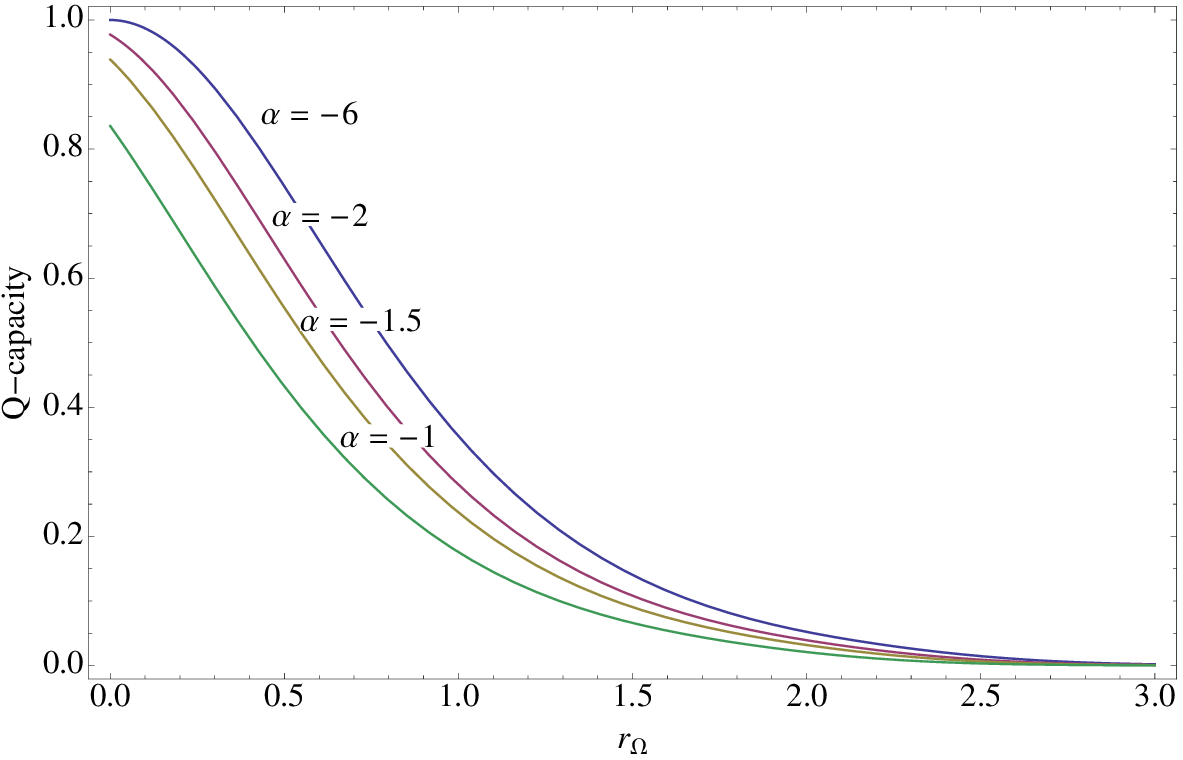}
\caption{The quantum capacity of an Unruh channel between Alice and Rob as the function of Hubble scale. The curves from top to bottom correspond to $\alpha=-6,-2,-1.5,-1$, respectively.}
\label{bosecapacity}
\end{figure}

The qudit Unruh channel $\mathcal{U}_d$ is the quantum channel defined by $\mathcal{U}_d(\rho_{A'})=\mbox{Tr}_{C}(U^\alpha\rho_{A'C'}(U^\alpha)^\dag)$, where $U^\alpha=\bigotimes_{i=1}^dU^\alpha_{A_iC_i}$. Here the modes $C$ beyond cosmological horizon are inaccessible for Rob therefore should be traced over. Since the qudit Unruh channel is conjugate degradable, the quantum capacity could be given by the optimized coherent information \cite{channel3}. Moreover, since the Unruh channel is covariant and conjugate degradable, its quantum capacity could be achieved with a maximally mixed input qudit $\pi_{A'}=\mathbb{I}_{A'}/d$. With reparameterization (\ref{uopa}) in mind, we can directly write down the quantum capacity with general $\alpha$ choice for $\mathcal{U}_d(\pi_{A'})$ as \cite{channel2}
\be
Q(\mathcal{U}_d)=\frac{1}{d}\bigg(\frac{1-e^{2\alpha}}{(\cosh r+e^\alpha\sinh r)^2}\bigg)^{d+1}\sum_{k=1}^\infty k\left(\begin{array}{c}d+k-1 \\k\end{array}\right)
\log\frac{d+k-1}{k}\bigg(\frac{\tanh r_\Omega+e^\alpha}{1+e^\alpha\tanh r_\Omega}\bigg)^{2(k-1)}
\ee
For particular example with $d=2$, this reduces to
\be
Q(\mathcal{U}_2)=\frac{1}{2}\bigg(\frac{1-e^{2\alpha}}{(\cosh r+e^\alpha\sinh r)^2}\bigg)^{3}\sum_{k=1}^\infty k(k+1)\log(1+1/k)\bigg(\frac{\tanh r_\Omega+e^\alpha}{1+e^\alpha\tanh r_\Omega}\bigg)^{2(k-1)}
\ee
which is plotted in Fig. \ref{bosecapacity} and show that quantum capacity of Unruh channel for various $\alpha$ always approach zero in the limit of infinite curvature. Since the only convergent point of bosonic entanglement negativity is $H_c=\infty$, we conclude that $H_c$ indicates zeros of quantum capacity of Unruh channel.

\subsection{Grassmann channel}

We now turn to fermionic quantum channel in de Sitter space which is more subtle due to Fermi-Dirac statistics of fermionic operators. For the fermionic state, a particular class of channels called Grassmann channels have been introduced in Minkowski space \cite{channel1}. In de Sitter space, we employed such fermionic quantum channel between comoving observer Alice and static observer Rob to transmit quantum information. 

Similar as Unruh channel, we should introduce an unitary operator consistent with $\alpha$-vacua. Since the Grassmann channel in Bunch-Davies choice resembles those in Minkowski space, the fermionic MA transformation (\ref{fermialpha}) as squeezing transformation induces a reparameterization on the unitary operator defined in \cite{channel1}, which gives
\be
U^\alpha_{AC}(\tilde{r}_\Omega)=\exp\bigg[\mbox{arctan}\bigg(\frac{\tan \tilde{r}_\Omega+e^\alpha}{1-e^\alpha\tan  \tilde{r}_\Omega}\bigg)(a^\dag c^\dag-ac)\bigg]
\label{uopaf}
\ee
where $A$ and $C$ denote right and left Rindler wedge as before, and fermionic operator $a$ satisfies canonical anticommunication relation $\{a_k,a_k^\dag\}=\delta(k-k')$ (similar relation hold for $c$). To see this fermionic unitary operator do induces correct Bogoliubov transformation, by using identity similar with (\ref{id})
\bea
U^\alpha_{AC}(\tilde{r}_\Omega)&=&\big(1+\tan^2\tilde{r}_\Omega\tilde{\Delta}^2\big)^{-\frac{1}{2}}\exp\bigg[\frac{\tan \tilde{r}_\Omega+e^\alpha}{1-e^\alpha\tan  \tilde{r}_\Omega}a^\dag c^\dag\bigg]\no\\
&&\times\exp\bigg[\frac{1}{2}\ln\big(1+\tan^2\tilde{r}_\Omega\tilde{\Delta}^2\big)(a^\dag a+ c^\dag c)\bigg]\exp\bigg[\frac{\tan \tilde{r}_\Omega+e^\alpha}{1-e^\alpha\tan  \tilde{r}_\Omega}a  c \bigg]
\label{idf}
\eea
we calculate the action of $d$ copies of the operator as
\bea
|\psi\ra_{AC} &=&\bigotimes_{i=1}^dU^\alpha_{A_iC_i}|0\ra_A|0\ra_C\no\\
&=&(1+\tan^2\tilde{r}_\Omega\tilde{\Delta}^2)^{-d/2}\sum_{k=0}^d(-)^{k(k-1)/2}\bigg(\frac{\tan \tilde{r}_\Omega+e^\alpha}{1-e^\alpha\tan  \tilde{r}_\Omega}\bigg)^k\sum_{n_1,\cdots,n_d}^{\tiny\left(\begin{array}{c}d \\k\end{array}\right)}|\vec{n}\ra_A|\vec{n}\ra_C
\eea
where $|\vec{n}\ra\equiv|n_1,\cdots,n_d\ra$ with $n_i\in\{0,1\}$. For particular example with $d=2$, it becomes
\[
|\psi\ra_{AC}=\frac{(\cos\tilde{r}-e^\alpha\sin\tilde{r})^2}{1+e^{2\alpha}}[|00\ra_A|00\ra_C+\tan\tilde{r}\tilde{\Delta}(|10\ra_A|10\ra_C+|01\ra_A|01\ra_C)-\tan^2\tilde{r}\tilde{\Delta}^2|11\ra_A|11\ra_C]
\]
which is consistent with (\ref{unruh-f-0}).

A fermionic qudit with multi-rail encoding $|\Psi\ra_{A'C'}=\sum_{i=1}^d\beta_ia_i^\dag|0\ra_{A'}|0\ra_{C'}=\sum_{i=1}^d\beta_i|i\ra_{A'}|0\ra_{C'}$ could be transmitted by $d$ copies of (\ref{uopaf}) to the final output state
\bea
|\Phi\ra_{AC} &=&\bigotimes_{i=1}^dU^\alpha_{A_iC_i}|\Psi\ra_{A'C'}\no\\
&=&(1+\tan^2\tilde{r}_\Omega\tilde{\Delta}^2)^{-(d-1)/2}\bigg(\sum_{i=1}^d\beta_ia_i^\dag\bigg)\sum_{k=1}^{d+1}(-)^{(k-1)(k-2)/2}\bigg(\frac{\tan \tilde{r}_\Omega+e^\alpha}{1-e^\alpha\tan  \tilde{r}_\Omega}\bigg)^{k-1}\no\\
&&\times\bigg[\sum_{N_i}|\cdots n_j\cdots\ra_A|\cdots n_j\cdots\ra_C\bigg]\no\\
&=&(1+\tan^2\tilde{r}_\Omega\tilde{\Delta}^2)^{-(d-1)/2}\sum_{k=1}^d(-)^{(k-1)(k-2)/2}\bigg(\frac{\tan \tilde{r}_\Omega+e^\alpha}{1-e^\alpha\tan  \tilde{r}_\Omega}\bigg)^{k-1}\no\\
&&\times\bigg[\sum_{N_i}\sum_{i\in I}^{d-k+1}\beta_i|\cdots n_j+1_j\cdots\ra_A|\cdots n_j\cdots\ra_C\bigg]
\label{uopfem}
\eea
where the second summation is over the set $N_k=\{(n_1,\cdots,n_d)|\sum_{j=1}^dn_j=k-1\}$ all possible states with $k-1$ fermions. For fixed $N_k$, the third summation is over $I=\{i|n_i=0\}$, and finally the multiplicative sign $(\pm)_i$ is dependent on the fermionic state of the $A$ subsystem \cite{channel1}.

\begin{figure}[hbtp]
\includegraphics[width=.7\textwidth]{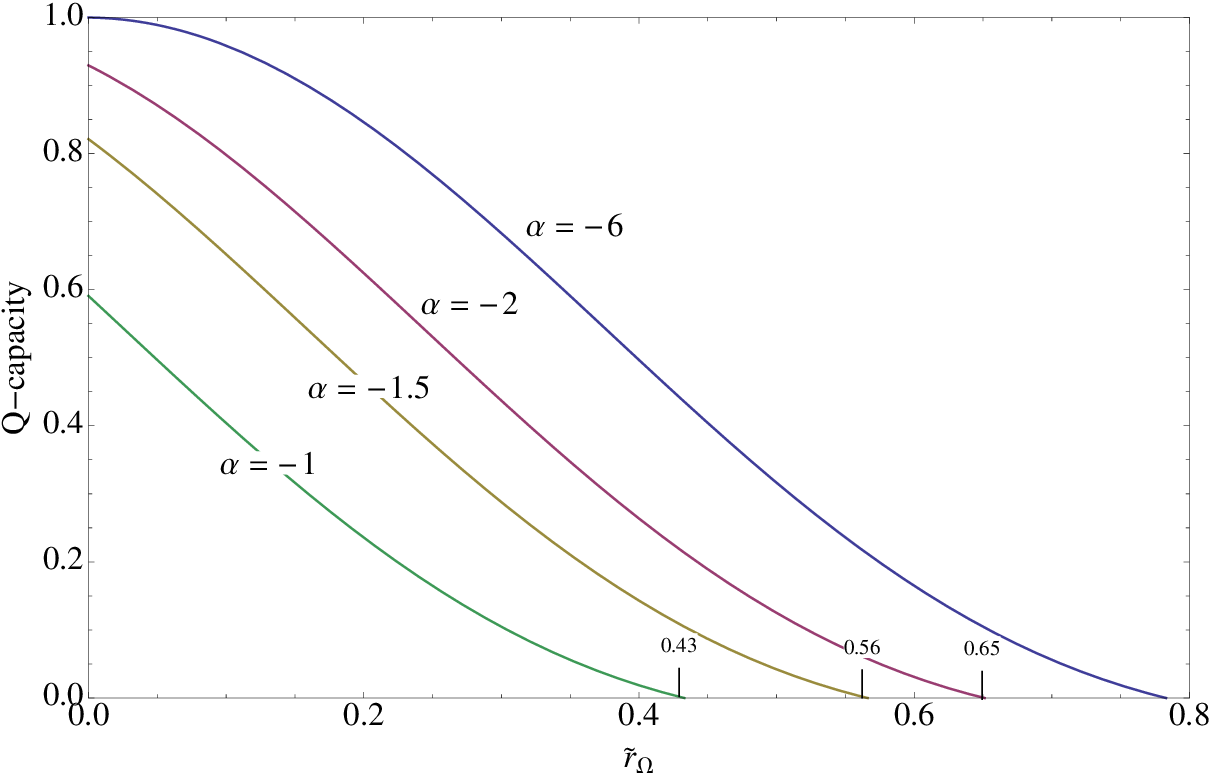}
\caption{The quantum capacity of a Grassmann channel between Alice and Rob as the function of Hubble scale. The curves from top to bottom correspond to $\alpha=-6,-2,-1.5,-1$, respectively. The zeros of quantum capacity have been indicated which are same convergent points $H_c$ of fermionic negativity.}
\label{fermicapacity}
\end{figure}

Now, the $d$-dimensional Grassmann channel $\mathcal{G}_d:\mathcal{B}(\mathbb{C}^d)\rightarrow\mathcal{B}(\mathbb{C}^{2^d-1}) $ is consisted of unitary transformation (\ref{uopfem}) along with tracing over the inaccessible modes beyond Rob's cosmological horizon, i.e., $\mathcal{G}_d(\rho_{A'})=\mbox{Tr}_{C}(U^\alpha\rho_{A'C'}(U^\alpha)^\dag)$. Similar as Unruh channel, the Grassmann channel is covariant and degradable, therefore the supremum of coherent information in (\ref{qcap}) can also be achieved by a maximally mixed input qudit. The quantum capacity formula in \cite{channel1} now reads for fermionic $\alpha$-vacua as
\be
Q(\mathcal{G}_d)=\frac{1}{d}\bigg[\frac{(\cos\tilde{r}-e^\alpha\sin\tilde{r})^2}{1+e^{2\alpha}}\bigg]^{d-1}\sum_{k=1}^d k\left(\begin{array}{c}d \\k\end{array}\right)\log k\bigg[\bigg(\frac{\tan \tilde{r}_\Omega+e^\alpha}{1-e^\alpha\tan  \tilde{r}_\Omega}\bigg)^{2(d-k)}-\bigg(\frac{\tan \tilde{r}_\Omega+e^\alpha}{1-e^\alpha\tan  \tilde{r}_\Omega}\bigg)^{2(k-1)}\bigg]
\ee
For $d=2$, this reduces to
\be
Q(\mathcal{G}_2)=\bigg[\frac{(\cos\tilde{r}-e^\alpha\sin\tilde{r})^2}{1+e^{2\alpha}}\bigg]^{3}\bigg[1-\bigg(\frac{\tan \tilde{r}_\Omega+e^\alpha}{1-e^\alpha\tan  \tilde{r}_\Omega}\bigg)^2\bigg]
\label{locus}
\ee
which is plotted in Fig. \ref{fermicapacity}. We observe that quantum capacity of the Grassmann channel becomes vanish at point $\tilde{r}_\Omega=0.43,0.56,0.65$ for $\alpha=-1,-1.5,-2$, exactly the same critical point $H_c$ appearing in (\ref{critical}). Therefore, we conclude that for fermionic field there is an one-to-one correspondence between $H_c$ and zeros of quantum capacity of Grassmann channel. Combining with the result in previous section, we finally arrive the conclusion that the link between the convergent point $H_c$ of entanglement and the zeros of quantum capacity is \emph{universal} for free quantum fields in de Sitter space.

\begin{figure}[hbtp]
\includegraphics[width=.7\textwidth]{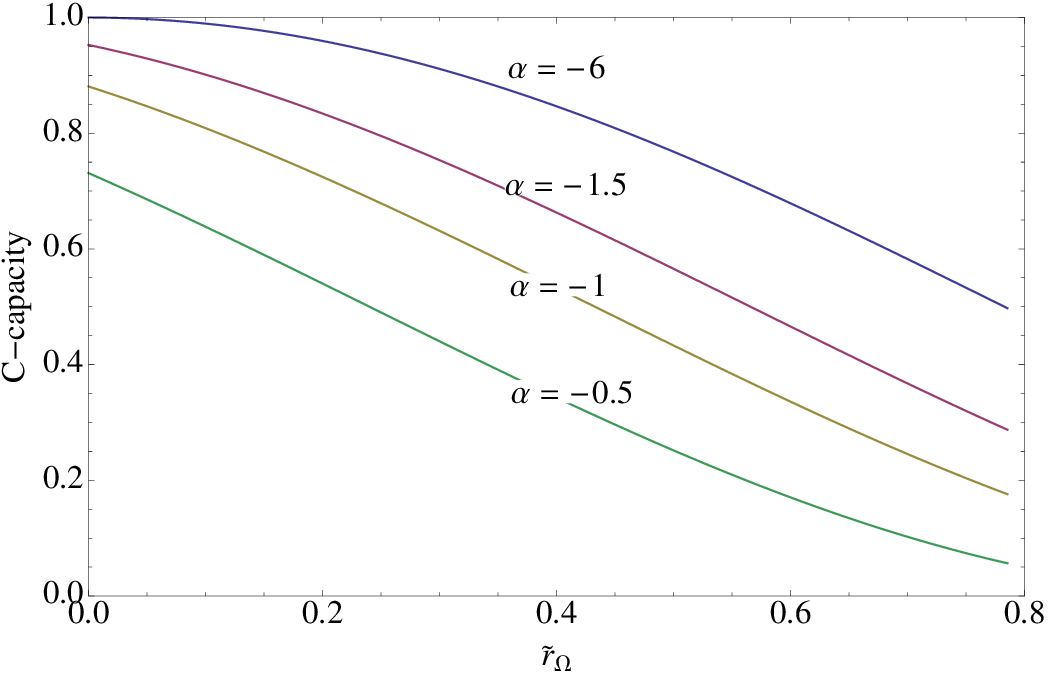}
\caption{The classical capacity of a Grassmann channel between Alice and Rob as the function of Hubble scale. The curves from top to bottom correspond to $\alpha=-6,-1.5,-1,-0.5$, respectively. The classical capacity is always nonnegative which means classical information could always be transmitted by Grassmann channel.}
\label{fermiccapacity}
\end{figure}

Finally, we demonstrate the pure quantum character of the convergent points $H_c$ of entanglement by showing that the classical capacity of Grassmann channel is always nonnegative. For general $\alpha$ choice, exploiting the result from \cite{channel1} that all Grassmann channels have the Holevo capacity additive, the classical capacity formula (\ref{ccap}) reads
\be
C(\mathcal{G}_d)=\log d-\bigg[\frac{(\cos\tilde{r}-e^\alpha\sin\tilde{r})^2}{1+e^{2\alpha}}\bigg]^{d-1}\sum_{k=1}^d\bigg(\frac{\tan \tilde{r}_\Omega+e^\alpha}{1-e^\alpha\tan  \tilde{r}_\Omega}\bigg)^{2(k-1)}\left(\begin{array}{c}d-1 \\k-1\end{array}\right)
\log k
\ee
For $d=2$, this reduces to
\be
C(\mathcal{G}_2)=\frac{(\cos\tilde{r}-e^\alpha\sin\tilde{r})^2}{1+e^{2\alpha}}
\ee
which is plotted in Fig. \ref{fermiccapacity}. One can learn that nothing special can happen at convergent points $H_c$. For Bunch-Davies choice, there is a residual amount of $C(\mathcal{G}_2)$ even at the limit $H\rightarrow\infty$, which means that classical correlation could be transmitted even with infinite spacetime curvature. For general $\alpha$, the residual classical capacity is more suppressed, in particular, $C(\mathcal{G}_2)\rightarrow0$ at the limit $H\rightarrow\infty$ if   $\alpha$ is approach $0^-$.
 
In summary, we show that for free quantum field modes in de Sitter space there is an one-to-one correspondence between the convergent points of quantum entanglement and the zeros of quantum capacity. In particular, there is a series of convergent points $H_c$ determined by fermionic $\alpha$-vacua (\ref{locus}) where no quantum information could be transmitted via a Grassmann channel. The fermionic entanglement is nonvanishing and divergent beyond $H_c$ as illustrated in Fig. \ref{ferminegativity}. Nevertheless, our result implies that this part of fermionic entanglement between Alice and Rob is not interesting, since it has no operational meaning in de Sitter space, while usually quantum entanglement should play as physical resource in real quantum information process. 

For instance, we sketched a typical quantum channel, i.e., quantum teleportation protocol in de Sitter space in Ref. \cite{FENG1}. To establish a teleportation protocol, which means to teleport an unknown state $|\psi\ra=\alpha|0\ra+\beta|1\ra$ from comoving detector Alice to a static detector Rob, a two-qubit Bell state should be shared by two observers. After Alice makes a joint projective measurement on her two logical qubits, the results of measurement could be sent from Alice for Rob who can then recover the unknown state by applying a proper unitary transformation on his qubit and complete the protocol in his local frame. Such a teleportation process in de Sitter space should be indeed influenced by the local Gibbons-Hawking radiation detected by Rob. Moreover, additional modification from the existence of minimal fundamental scale should also be taken into account if one starts from general $\alpha$-vacua rather than the Bunch-Davies choice. As shown in Ref. \cite{FENG1}, one can compare different choice of de Sitter vacua by the fidelity of transferred state, which has intrinsic relation to quantum channel capacity \cite{NIEL}.

\section{Conclusion and discussion}

It has been known for a long time that quantum correlations, in particular quantum entanglement, not only
plays a key role in quantum information science, but can also provide powerful means for other
research areas such as condensed matter physics and cosmology \cite{cui,RQI14}. In this paper, we give a new interpretation this universality to the physics at some fundamental scales, like Planckian or stringy, by its influence on quantum correlations in certain quantum information tasks. We analyze the quantum correlations of free fields in de Sitter space, while the Planckian modifications presented by vacuum ambiguity has been considered. Beside the existence of degradation behavior of entanglement same as in previous RQI works \cite{RQI7,RQI8,RQI9,RQI10}, we exploit some new features appearing in de Sitter background. Firstly, we show that the correction from fundamental scale are encoded in patterns of degradation of quantum correlations for a static observer in de Sitter space. Comparing with the Bunch-Davies choice, the possible Planckian physics cause some extra decrement on the quantum correlation, which may provides the means to detect quantum gravitational effects via quantum information methodology. Secondly, by constructing proper Unruh modes respect $\alpha-$vacua beyond SMA, we show the convergent feature of negativity of both bosonic and fermionic entanglement. In particular, for fermionic entanglement, we find that the convergent points $H_c$ are dependent on the choice of $\alpha$. To explore the physical meaning of $H_c$, we construct Grassmann quantum channel between Alice and Rob in de Sitter space, and show that the quantum capacity of channel becomes vanishing at these convergent points, which means no quantum correlations could be transmitted in de Sitter space. On the other hand, for a bosonic Unruh channel in de Sitter space, we show that the quantum capacity of an Unruh channel is always positive and becomes vanishing at $H_c=\infty$. Therefore, ultimately, we prove that the one-to-one correspondence between $H_c$ and the zeros of quantum capacity is universal for both bosonic and fermionic field.

In this work, we employ the Bogoliubov transformation technology wildly used in RQI literature to analyze the global modes in spacetime. This method allow us to exploit the general $\alpha-$states quite straightforwardly as the MA transformation utilizing antipodal map over whole de Sitter manifold \cite{ALPHA3,ALPHA4,ALPHA6}. To consider a more physical scenario, it is more satisfactory to use local particle detectors approach as in \cite{RQI11, RQI12, YU}. However, to our knowledge, such detector picture (especially for general $\alpha$-state) in de Sitter space still suffer from its ill-defined renormalizability and stability. While there are some clues to this problem (with specific operators ordering in propagator) already \cite{nonlocal2}, it would be very interesting to investigate the RQI in this framework in future. 

At the present stage, we have considered a standard de Sitter spacetime with constant curvature, therefore the parameter $r_\Omega$ for bosonic field and $\tilde{r}_\Omega$ fro fermionic field depend only on the $|k|$ of fields and time-independent Hawking temperature $H/2\pi$ (cf. (\ref{u-s}) and (\ref{u-s-f})). In a real inflation model, the quantum correlation between field modes, especially between cosmological perturbation, suffer a more complicate decoherence. Nevertheless, the analysis present in this work could be straightly generalized to that case \cite{future1}, which may shed new light on the problem of quantum-to-classical transition of the perturbations. In particular, starting from different choice of initial state, the quantum communication in cosmological background should be influenced, indicated by the channel's capacity \cite{FENG1}. On the other hand, with respect to $\alpha$-vacua, backreaction acts to increase the de Sitter horizon and therefore makes de Sitter taller \cite{back1}. In an expanding universe, backreaction would impose constraints on short distance effects in the CMB \cite{back2}. Therefore when discuss the possible Planckian modification on quantum communication in universe, the constraints from backreaction should be taken into account. We will explore all these issues elsewhere in future.

The design of the intelligent RQI experiments is of course the next challenging open problem \cite{entgravity}. Recently developed quantum metrology for relativistic quantum fields \cite{metrology} may open a new window to directly detection of extremal sensitive Planckian effects. On the other hand, a more pragmatic approach is to simulate these Planckian modifications by analogue gravity experiments, like using ion trap. In detector picture \cite{ion}, replacing the conformal time $\eta$ with the experimenter's clock time, the detector should evolve with respect to the simulated proper time $t$ which is equal to the cosmic time. To simulate vacuum ambiguity, the detector's response function represents the probability for an ion excited by interaction with the field. For general $\alpha\neq-\infty$, the standard Wightman function would be modified \cite{ALPHA4}. Therefore, the ion analogue of this function $\la\phi_m(\xi)\phi_m(\xi')\ra$ should be evaluated in some motional-state\cite{ionsqueeze}, since the $\alpha$-vacua can be interpreted as squeezed states over Bunch-Davies vacuum state.

\section*{ACKNOWLEDGEMENT}

This work is supported by the Australian Research Council through DP110103434. H. F. acknowledges the support of NSFC, 973 program (Grant No. 2010CB922904). W.L.Y. acknowledges the support of NSFC (Grant Nos. 11075126 and 11031005). C.Y.S. acknowledges the support of NSFC (Grant Nos. 11147017).

\end{document}